\documentclass[iop]{emulateapj}
\usepackage{apjfonts}
\usepackage{natbib,graphicx,amssymb}

\shorttitle{Giant Planet and Brown Dwarf Mean Opacities}
\shortauthors{Freedman et al.}

\newcommand{\teff}{$T_{\rm eff}$}

\newcommand{\cp}{\citep}
\newcommand{\ct}{\citet}

\slugcomment{Draft for ApJS}

\begin{document}

\title{Gaseous Mean Opacities for Giant Planet and Ultracool Dwarf Atmospheres over a Range of Metallicities and Temperatures}

\author{Richard S. Freedman\altaffilmark{1}$^,$\altaffilmark{2}, Jacob Lustig-Yaeger\altaffilmark{3}$^,$\altaffilmark{4}, Jonathan J. Fortney\altaffilmark{4}, Roxana E. Lupu\altaffilmark{1}$^,$\altaffilmark{2}, Mark S. Marley \altaffilmark{2}, Katharina Lodders \altaffilmark{5}}

\altaffiltext{1}{SETI Institute, Mountain View, CA, USA; Richard.S.Freedman@nasa.gov}
\altaffiltext{2}{Space Science and Astrobiology Division, NASA Ames Research Center, Moffett Field, CA, USA}
\altaffiltext{3}{Department of Physics, University of California, Santa Cruz, CA 95064}
\altaffiltext{4}{Department of Astronomy and Astrophysics, University of California, Santa Cruz, CA 95064}
\altaffiltext{5}{Planetary Chemistry Laboratory, Washington University, St. Louis, MO, USA}

\begin{abstract}
We present new calculations of Rosseland and Planck gaseous mean opacities relevant to the atmospheres of giant planets and ultracool dwarfs.  Such calculations are used in modeling the atmospheres, interiors, formation, and evolution of these objects.  Our calculations are an expansion of those presented in Freedman et al.~(2008) to include lower pressures, finer temperature resolution, and also the higher metallicities most relevant for giant planet atmospheres.  Calculations span 1 $\mu$bar to 300 bar, and 75 K to 4000 K, in a nearly square grid.  Opacities at metallicities from solar to 50 times solar abundances are calculated.  We also provide an analytic fit to the Rosseland mean opacities over the grid in pressure, temperature, and metallicity.  In addition to computing mean opacities at these local temperatures, we also calculate them with weighting functions up to 7000 K, to simulate the mean opacities for incident stellar intensities, rather than locally thermally emitted intensities.  The chemical equilibrium calculations account for the settling of condensates in a gravitational field and are applicable to cloud-free giant planet and ultracool dwarf atmospheres, but not circumstellar disks.  We provide our extensive opacity tables for public use.
\end{abstract}

  
\section{Introduction}
A quantitative understanding of radiation transport in the cool molecule-dominated regions in planetary and ultracool dwarf atmospheres is essential to many aspects of understanding the temperature structure, thermal evolution, and formation of these objects.

In \ct{Freedman08} (hereafter, F08) we presented a detailed discussion of atomic and molecular line opacities used by our group in modeling the atmospheres of brown dwarfs and giant planets \cp[e.g.][]{Marley02,Fortney05,Saumon08,Fortney08a,Marley10}.  Since mean opacities can also be widely used in many contexts, in F08 we also computed Rosseland and Planck mean opacities from 75 to 4000 K, 0.3 mbar to 300 bar, at metallicities of [M/H] = -0.3, 0.0, and +0.3.  These tabulations have since found wide use in the communities working to understand giant planet formation \cp[e.g.,][]{Mordasini12a,Bodenheimer13}, the temperature structure of giant planet atmospheres \cp[e.g.][]{Paxton13}, and planetary and ultracool dwarf thermal evolution \cp[e.g.,][]{Batygin11,Valencia13,Paxton13}. 

Here we extend our previous work in a number of important aspects.  In addition to updates in opacities of particular molecules (described below), these new calculations are performed over a much larger phase space.  Mean opacity calculations at pressures from 1 $\mu$bar to 300 bars and 75 to 4000 K are presented, which is a much larger range in pressure, in particular at higher temperatures and lower pressures, compared to F08.  The overall temperature resolution of this compilation is also much finer.  Furthermore, we present calculations over a wide range of metallicities, from solar up to [M/H]=+1.7 ($\sim$50$\times$ solar), in several increments.  These high metallicities may well-approximate the metal-rich atmospheres of giant planets, up to levels of the solar system's ice giant planets, Uranus and Neptune \cp{Guillot07}.  Finally, for use in models of irradiated planetary atmospheres, we calculate mean opacities where the temperature in the weighting function is not the local temperature, but rather stellar blackbody temperatures from 3000-7000 K, to simulate mean ``incident flux'' or ``visible'' opacities to understand the absorption of incident stellar flux in planetary atmospheres.

\section{Sources of Opacity and Chemical Abundances}
Modern calculations of warm planetary atmospheres must draw from a variety of databases, including HITRAN \cp{Rothman:2009}, HITEMP \cp{Rothman:2010},  ExoMol \cp{Tennyson12}, and other sources.  \ct{Sharp07} and F08 provided excellent overviews of available opacities for giant planet and brown dwarf atmosphere modeling.  Much of this information remains current.  Our major changes since F08 are in ammonia, molecular hydrogen Collision-Induced Absorption (CIA), methane, and carbon dioxide.  For NH$_3$ the updated reference is \ct{Yurchenko:2011}, which is a first-principles calculation for temperatures up to 1500 K, which replaces our use of HITRAN.  For H$_2$ CIA we use \ct{Richard:2012} in place of previous work by \ct{Borysow02}.  In our recent work on brown dwarf model atmospheres, we have found these new opacity databases  provide better fits to observed brown dwarf spectra \cp{Saumon12,Morley12}.  

Two very recent opacity calculations for methane and carbon dioxide are now included.  For methane we make use of the new first-principles line lists of \ct{Yurchenko13} and \ct{Yurchenko14}, which replaces our previous use of older sources \cp{Brown:2005,Strong:1993,Wenger:1998}.  We also now include the opacity contributions of CO$_2$ \citep{Huang13,Huang14}, which is a very minor player at solar metallicity, but rises in importance in metal-rich atmospheres as its abundance grows quadratically in metallicity \cp{Lodders02}.  All of our relevant molecular opacities, as well as the references, are collected in Table 1. 

As in our previous tabulations, here we do not account for the opacity of liquids or solids, which in the relatively high gravity atmospheres of planets and brown dwarfs may be confined to cloud decks.  The opacity of these clouds decks is likely to be a strong function of poorly known parameters (vigor of vertical mixing, particle size, particle size distribution) and also other parameters that cannot be readily incorporated into tables for general application over a range of object surface gravities and temperature structures.  Therefore, to maximize the utility of these calculations in the wide variety of potential applications, we continue to exclude the opacity of condensates.  The condensation of cloud species, and their removal from the gas phase, is described below.

\subsection{Chemistry Calculations} 
The abundances of relevant atoms and molecules must be known at a given pressure and temperature in order to assess the wavelength dependent opacity.  As in F08 we make use of the chemistry calculations of K. Lodders and collaborators, using the base solar abundances of \ct{Lodders03}.  The chemistry calculations are described in a series of papers \cp[e.g.,][]{Fegley94,Lodders99,Lodders02,Lodders02b,Lodders06,Lodders09}, which use the CONDOR code.  Chemistry is calculated  assuming that thermodynamic equilibrium is reached.

This approximation is generally realized at high temperatures, where chemical reactions speed up dramatically.  However, it becomes less correct at lower temperatures, where the timescale for vertical mixing in an atmosphere can be faster than relevant chemical conversion timescales.  These effects have been explored in solar system giant planets, exoplanets, and brown dwarfs in many papers \cp[e.g.,][]{Prinn77,Fegley96,Saumon03,Saumon06,Hubeny07,Visscher11}. This can generally lead to an incorrect estimation of the mixing ratios of carbon-bearing and nitrogen-bearing molecules, as the molecules CO and N$_2$ often have very long timescales for conversion to other molecules at low temperature.  However, these effects depend strongly on the exact temperature structure of an atmosphere, as well the vertical mixing time.  These two effects cannot be generalized and incorporated with these tables, which are meant to be used over a wide phase space.  Perhaps most importantly, these mixing ratio differences are not a first-order effect.

As described in F08, the chemistry calculations are applicable to giant planets and ultracool dwarfs, which have significant self-gravity, unlike in protostellar disks.  In disks, condensates remain well-mixed and in contact with the surrounding gas, which allows for further chemical reactions between condensates and gases.  However, within planets and brown dwarfs, condensates form clouds that are typically vertically confined to a narrow region, often less than a gas scale height. Above a given refractory cloud, at lower pressures, the atmosphere is then gradually depleted in the compounds (atoms or molecules) that have gone into forming the condensate.  Simply, the availability of these refractory compounds in the gas becomes greatly diminished.  Whether or not the gas phase remains in contact with condensates can have dramatic effects on chemical equilibrium calculations \cp{Burrows99,Lodders99,Allard01,Lodders02} with decreasing temperature.  As discussed in F08,  all available evidence from brown dwarfs and the solar system's giant planets points towards this method of ``condensation, then depletion" of chemistry calculation as being most correct for these objects.

\subsection{Wavelength-Dependent Opacities}
\begin{figure}[ht] 
\begin{center}
 \includegraphics[width=2.70in,angle=90]{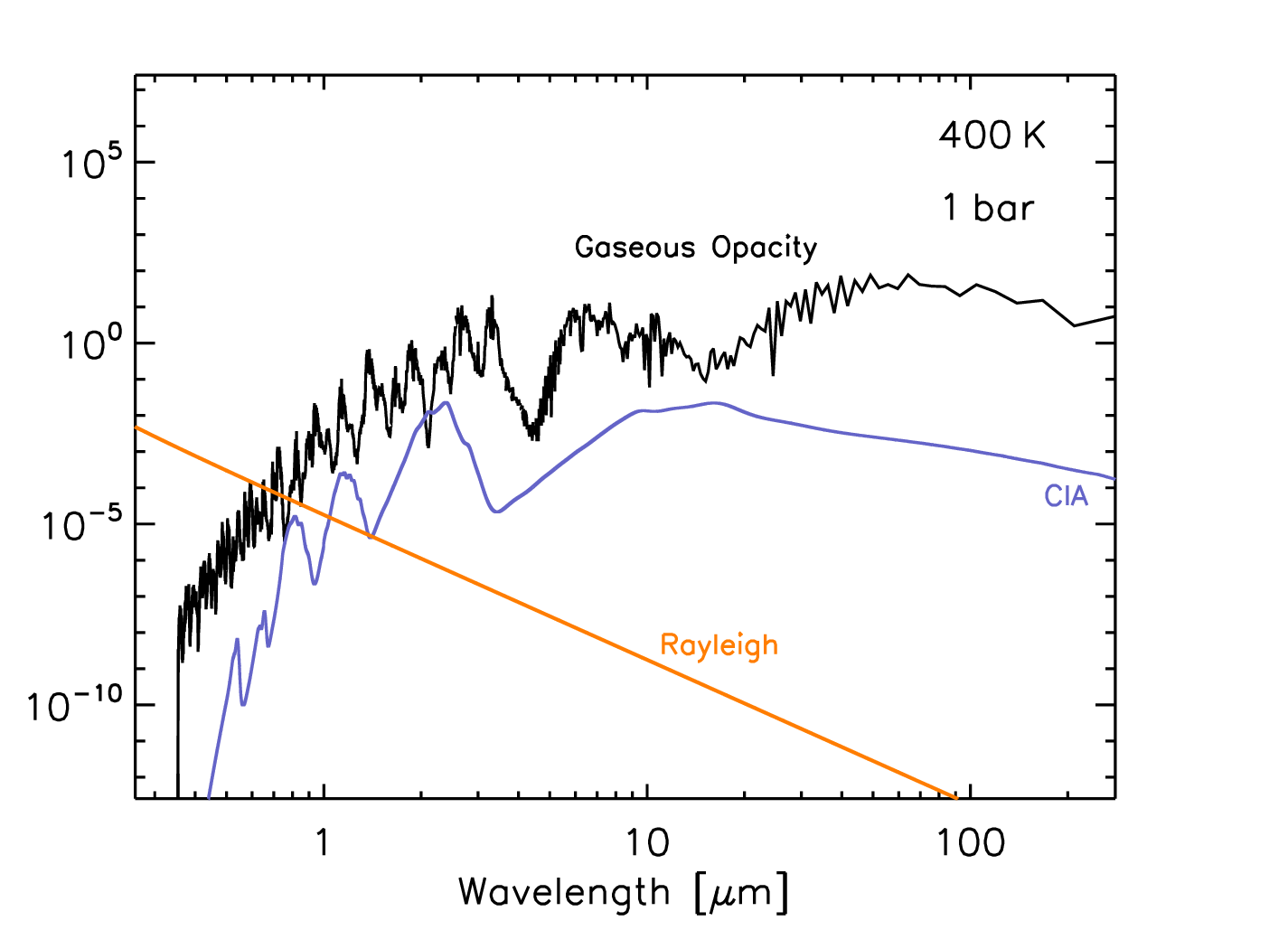}
 \includegraphics[width=2.70in,angle=90]{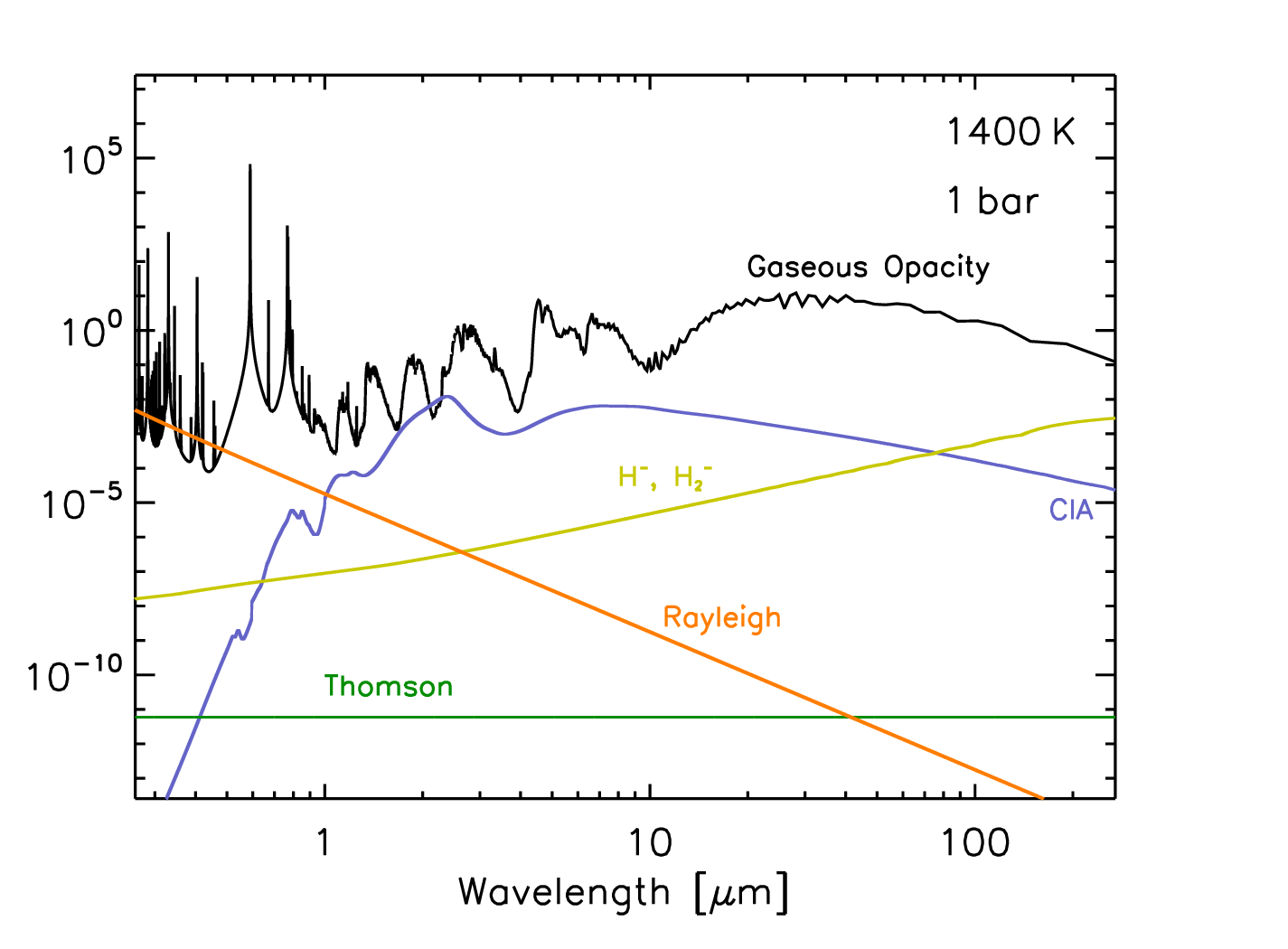}
 \includegraphics[width=2.70in,angle=90]{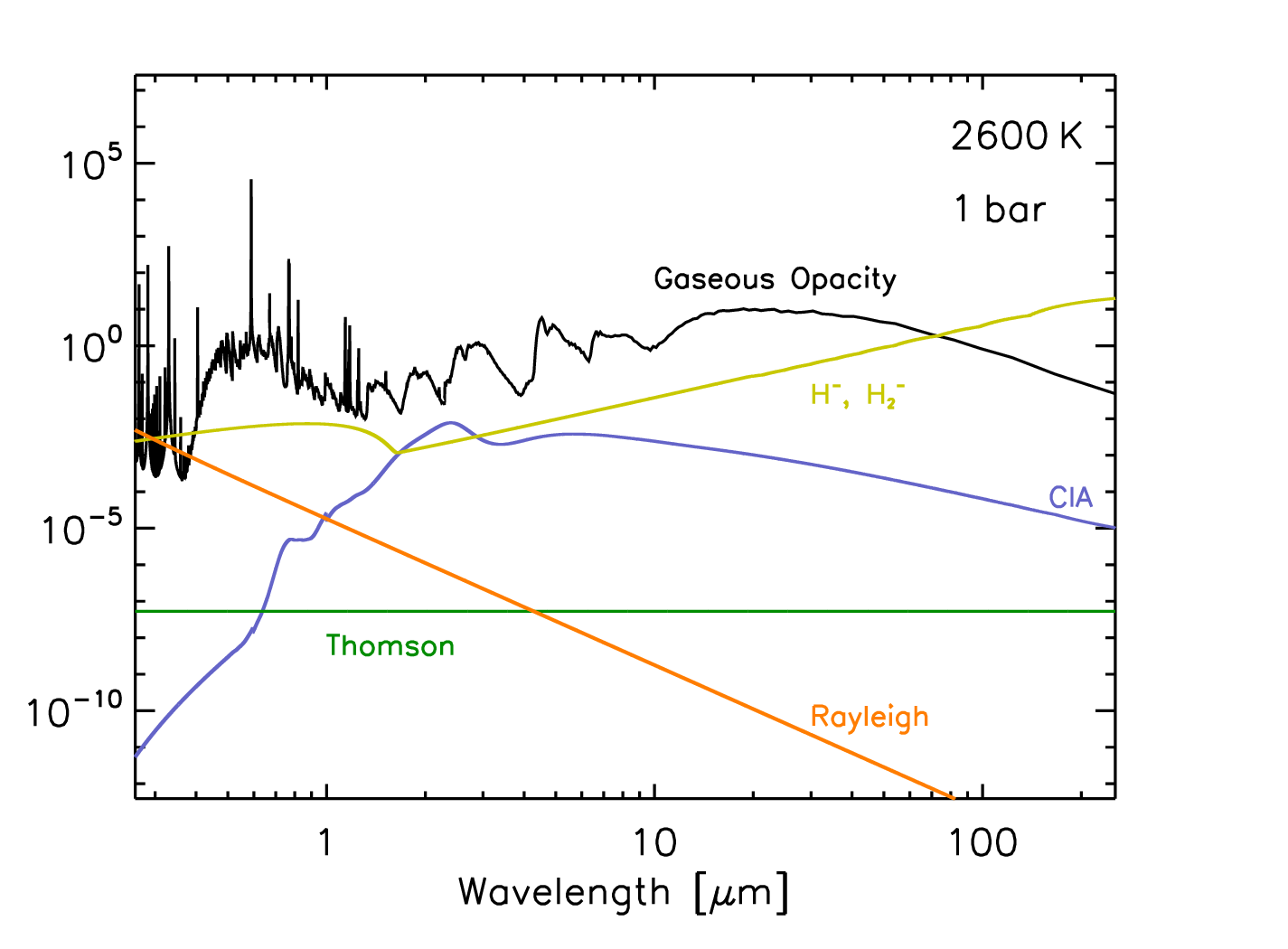}
 \caption{Relative contributions to the total gaseous opacity at 1 bar at 400 K (top), 1400 K (middle) and 2600 K (bottom).  Thomson scattering is in green, Rayleigh scattering is in orange, H$_2$ collision induced absorption (CIA) is blue, negative hydrogen ions (H$^{-}$, H$_2$$^{-}$) are shown in yellow, and opacity of all neutral atoms and molecules are shown in black.}
   \label{example}
\end{center}
\end{figure}

Our Wavelength-dependent opacities are calculated in wavenumber space over a grid corresponding to wavelengths from 0.268 to 227 $\mu$m.  In Figure \ref{example} we show the wavelength dependent opacity at 400 K, 1400 K, and 2600 K, all at 1 bar.  The plots are broken up to show the relative contributions of a few particular sources of opacity.  In black is shown the dominant opacity source at all temperatures, which is that of neutral atoms and molecules.  Water vapor is present at all of these temperatures, and dominates in the infrared.  In the optical at 400 K, there is relatively less opacity, but at warmer temperatures neutral atomic alkalis are present, which provide important opacities at 1400 K.  At 2600 K, TiO and VO gases are also important and generally overwhelm the alkali opacity.

Rayleigh scattering is generally important at the shortest wavelengths at all temperatures (orange), as is the CIA opacity of H$_2$ molecules (blue).  The contributions of electrons (Thomson scattering, in green) and negative hydrogen ions (yellow) are negligible at low temperatures but rise in importance at the highest temperatures as more free electrons are available.
\subsection{Mean Opacities}
The wavelength dependent opacities are tabulated at 1060 pressure-temperature \emph{P--T} points, on a nearly square grid.  The high pressure and low temperature corner of phase space is not included ($\sim$ 100 K and 100 bar), as these conditions do not occur in atmospheres.  For reference, the F08 calculations were at 324 points, and did not go to pressures below 0.3 mbar.  In addition, high temperatures were only included at high pressures, while our revised grid is nearly square. The grid is sampled in temperatures from 75--4000 K and finely sampled from 100-350 K (10 K between points) where water and ammonia condensation change the mixing ratio of these opacity sources quite dramatically.  We compute the Rosseland Mean (RM) opacity, defined as:
\begin{equation}
\frac{1}{\kappa_{_{\text{RM}}}} = \frac{\displaystyle \int_{0}^{\infty} \frac{1}{\kappa_\lambda} \frac{dB_\lambda}{dT} d\lambda}{\displaystyle \int_{0}^{\infty} \frac{dB_\lambda}{dT} d\lambda} 
\label{RM}
\end{equation} 
and the Planck Mean (PM) opacity, defined as:
\begin{equation}
\kappa_{_{\text{PM}}} = \frac{\displaystyle \int_{0}^{\infty} \kappa_\lambda B_\lambda d\lambda}{\displaystyle \int_{0}^{\infty} B_\lambda d\lambda},
\label{PM}
\end{equation}
where $\kappa_\lambda$ is the wavelength-dependent opacity and $B_\lambda$ is the Planck function, at all of these 1060 \emph{P--T} points.

\begin{figure}[ht] 
\begin{center}
 \includegraphics[width=2.70in,angle=90]{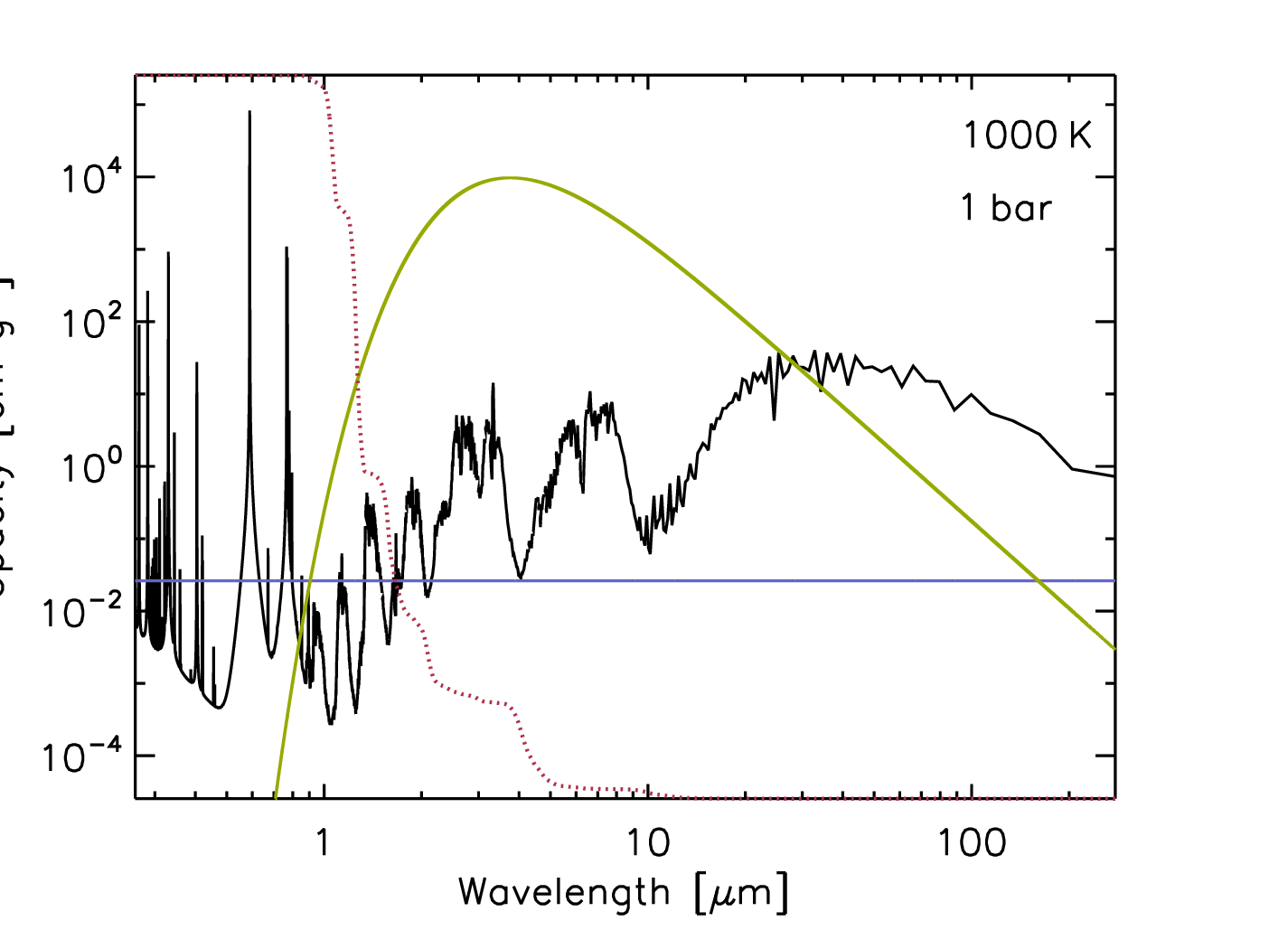}
 \includegraphics[width=2.70in,angle=90]{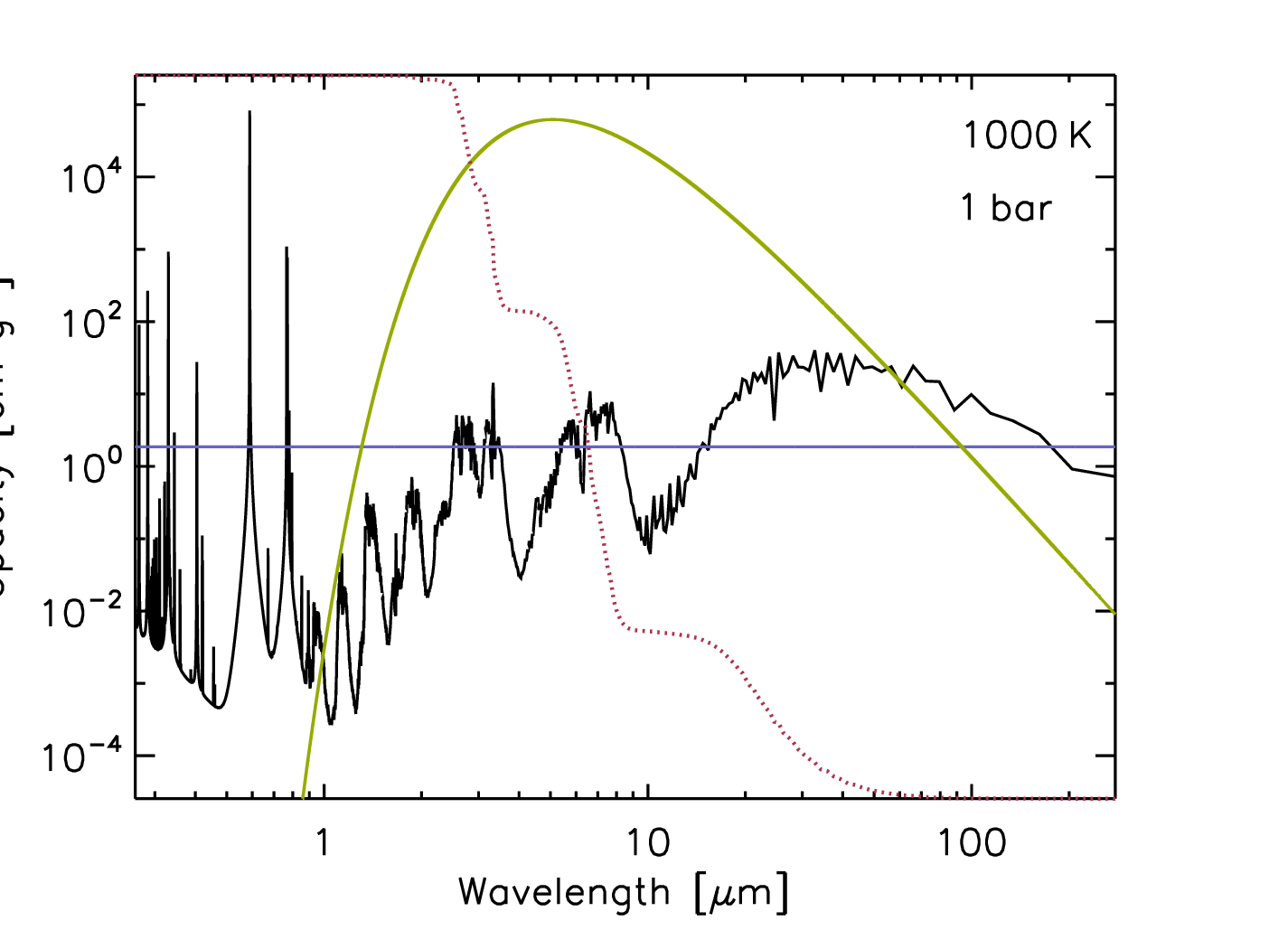}
 \caption{Sample calculation of the RM (top) and PM (bottom) at 1000 K and 1 bar.  The total gaseous opacity is shown in black, the weighting functions are shown in green, and the mean opacity is shown in blue.  The running opacity total from right to left, shown in dotted red, is shown on a \emph{linear} y-axis from bottom to top.  This clearly shows opacity windows contributing to the RM and strong bands contributing to the PM.}
   \label{RMPM}
\end{center}
\end{figure}

As an illustration, in Figure \ref{RMPM} at 1 bar and 1000 K, we show in black the total wavelength-dependent opacity, with the weighting function for RM opacity (top) and PM opacity (bottom).  The peak of both weighting functions is at the same wavelength, but they have a different functional form and shape.  In blue are the calculated RM and PM values.  For reference the running integral of the mean opacities are shown in dotted red on an arbitrary linear axis, starting from long wavelengths.  As expected, the dominant contributions to the RM opacities are in opacity minima, while for the PM opacities, these occur at opacity maxima.  It is interesting to note that, as has been pointed out as least as far back as \ct{King56} \cp[see also][for a modern update]{Parmentier14a}, that the ratio of the RM to the PM in an atmosphere is a measure of how important wavelength-dependnent (non-gray) effects are in determining the atmospheric temperature structure.

Figure \ref{RM} shows the RM opacity calculation on the entire grid, presented as a colored contour plot.  The general trend is one of the gradual loss of gaseous opacity with lower temperatures.
\begin{figure}
\centering 
%\hspace{3in}
\includegraphics[scale=0.36,angle=90]{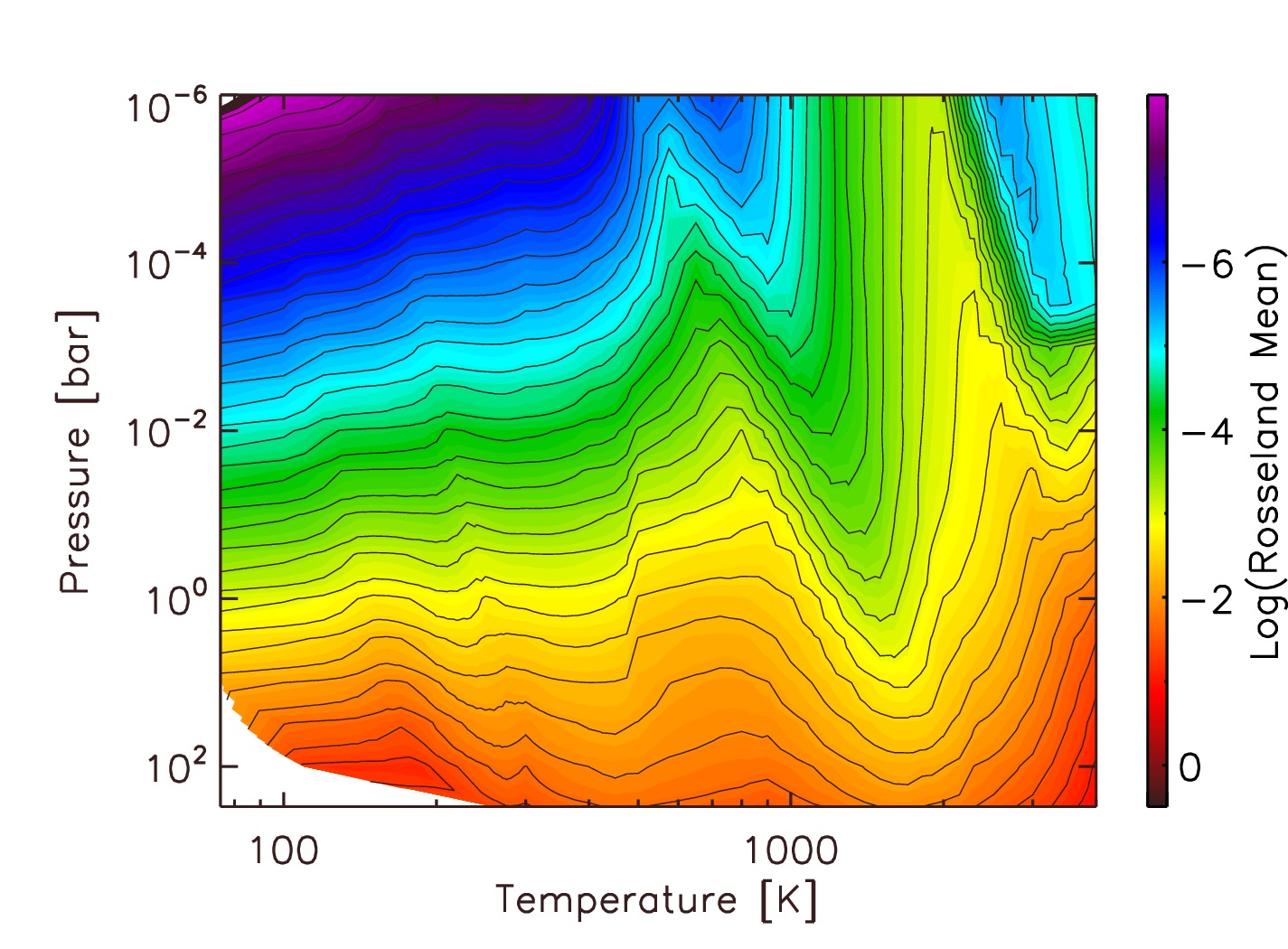}
\caption{Local Rosseland Mean opacities, at solar metallicity, over the grid of temperature and pressure calculated in this work.  The opacity units are cm$^2$ g$^{-1}$.
\label{RM}}
\end{figure}

\begin{figure}
\centering
%\hspace{3in}
\includegraphics*[scale=0.36,angle=90]{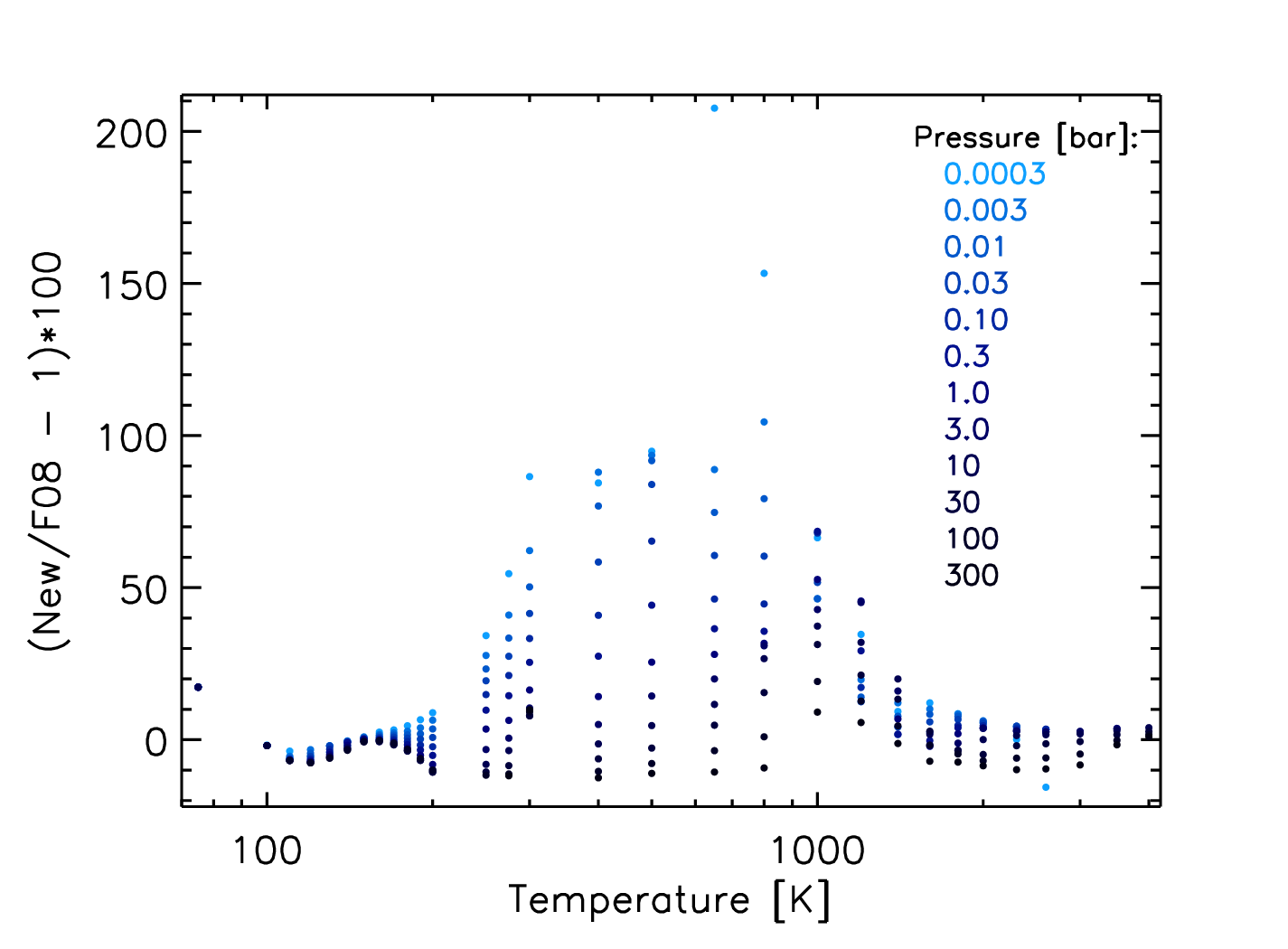}
\caption{Percent deviation of new RM opacities from those calculated by F08. Percent deviation is plotted versus local temperature, where pressure is denoted by the shade of blue--the darker the blue, the higher the pressure.  Differences are generally within 10\% over most of the temperature and pressure range.  Larger differences from 200 - 1000 K are generally due to the updated NH$_3$ opacity database.
\label{F08compare1}}
\end{figure}

Over the \emph{T--P} range where the F08 and our present calculations overlap (324 points at $P>0.3$ mbar) we can readily compare these calculations to the previous generation.  This is shown for the RM opacities in Figure \ref{F08compare1}, where pressure is color-coded.  In general, the agreement is at the level of 10\% or better.  H$_2$ CIA is important over the entire temperature range, and we find that $\sim$80\% of the generally small differences between the two calculations are due to this upgrade to the opacities from \ct{Richard:2012}.  Larger differences (though never more than a factor of 2) are seen below $\sim$ 1000 K, where warm ammonia is the main nitrogen carrier.  The first-princples \ct{Yurchenko:2011} line list is a significant improvement over our previously-used HITRAN database, which lacked proper temperature dependence and was quite incomplete in the near infrared.  As ammonia is lost into a cloud below $\sim$200 K, these differences between the F08 and new calculations are minimized.

\begin{figure}
\centering
%\hspace{3in}
\includegraphics*[scale=0.36,angle=90]{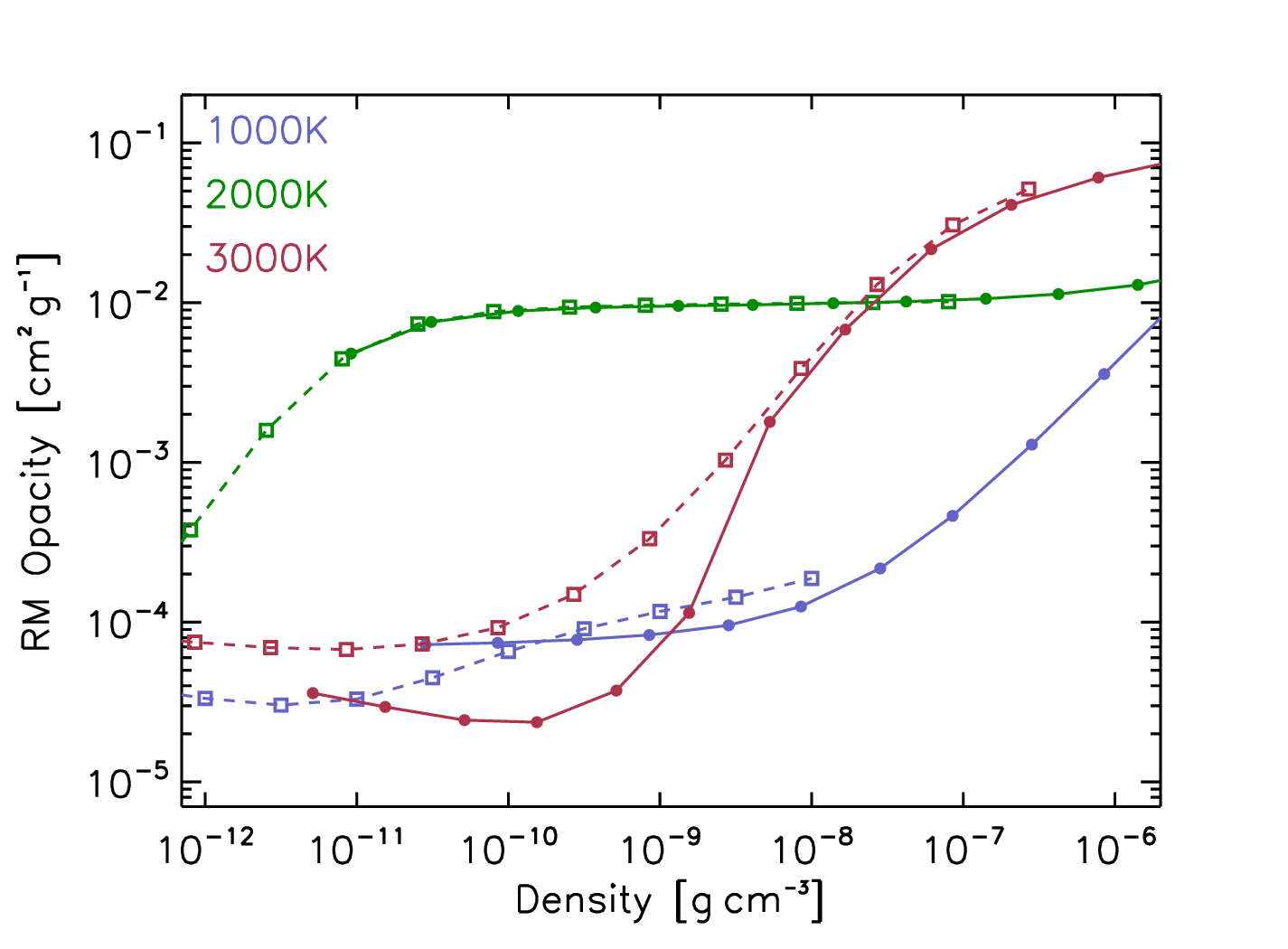}
\caption{Rosseland mean opacities for our calculations (solid lines with filled circles) compared to a grain-free $X=0.70$, $Z=0.02$ calculation by \ct{Ferguson05}, table ``cunha06.nog.7.02.tron'' (dashed lines with open squares).  The agreement appears reasonably good in the range of overlap, particularly at 2000 K.
\label{wich}}
\end{figure}

We can also compare our calculations to those of another group.  In Figure \ref{wich} we compare our RM results at 1000, 2000, and 3000 K to those of \ct{Ferguson05}.  The chosen table, grain-free ``cunha06.nog.7.02.tron'' has a metallicity ($X=0.70$,$Z=0.02$) somewhat similar to our own solar composition.  The \ct{Ferguson05} calculations only have moderate density overlap with our own, since their calculations are most appropriate for low-density disks while ours are for denser planetary atmospheres.  However, the agreement does appear reasonably good.

\section{Effects of Metallicity}
While the original F08 opacity calculations spanned a fairly narrow range of metallicities, from [M/H]=-0.5 to +0.5, here we expand these calculations to include [M/H] = 0, +0.5, +0.7, +1.0, +1.5, and +1.7, roughly corresponding to 1$\times$, 3$\times$, 5$\times$, 10$\times$, 30$\times$, and 50$\times$ solar abundances.  For reference, Jupiter's atmosphere is approximately 3-5$\times$ solar \cp{Wong04}, Saturn's atmosphere is enhanced in methane by a factor of $\sim$~10 \cp{Flasar05}, and Uranus and Neptune in methane by a factor to $\sim$50 \cp{Guillot07}.

\begin{figure}[ht] 
\begin{center}
 \includegraphics[width=2.70in,angle=90]{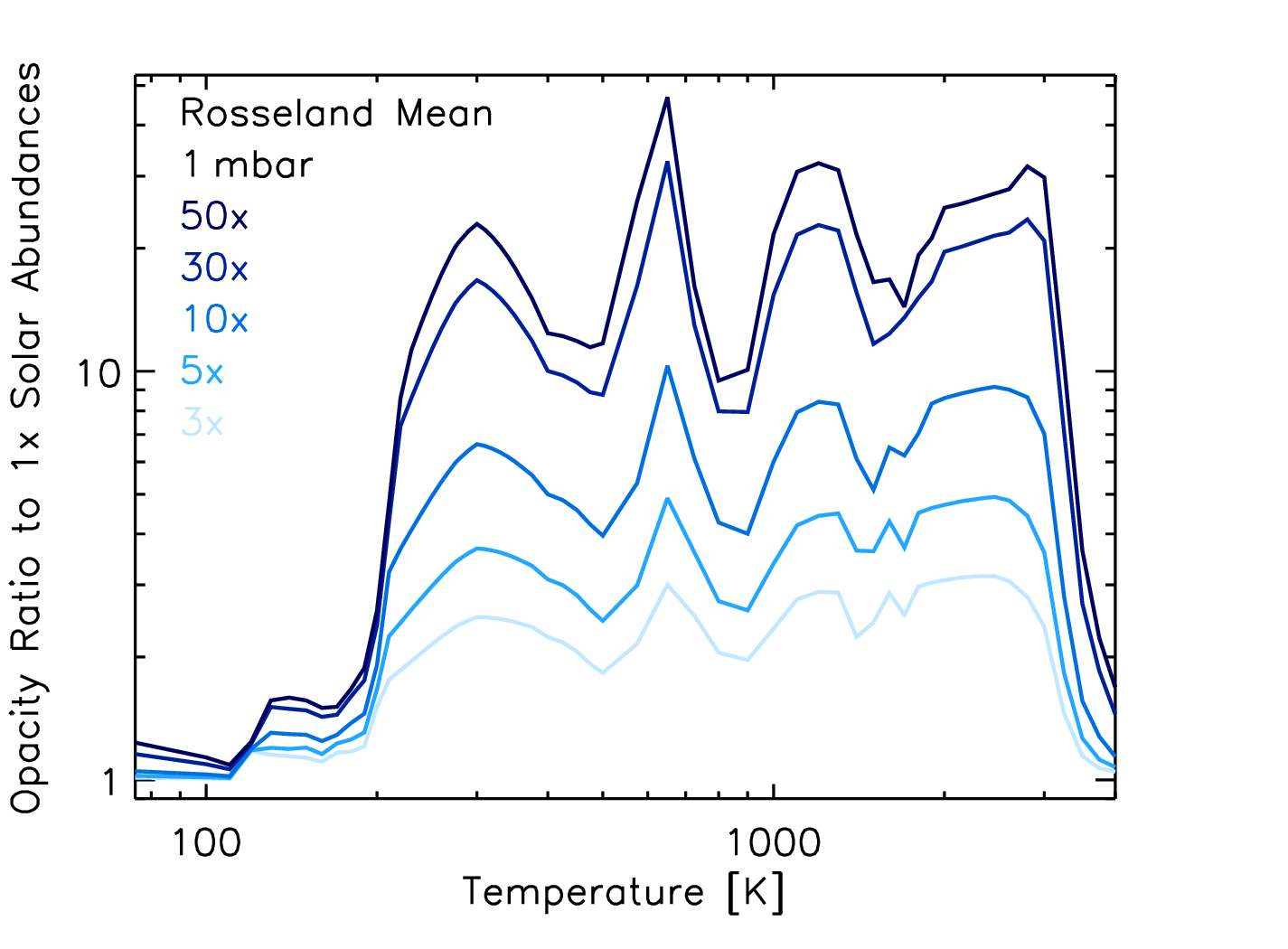}
 \includegraphics[width=2.70in,angle=90]{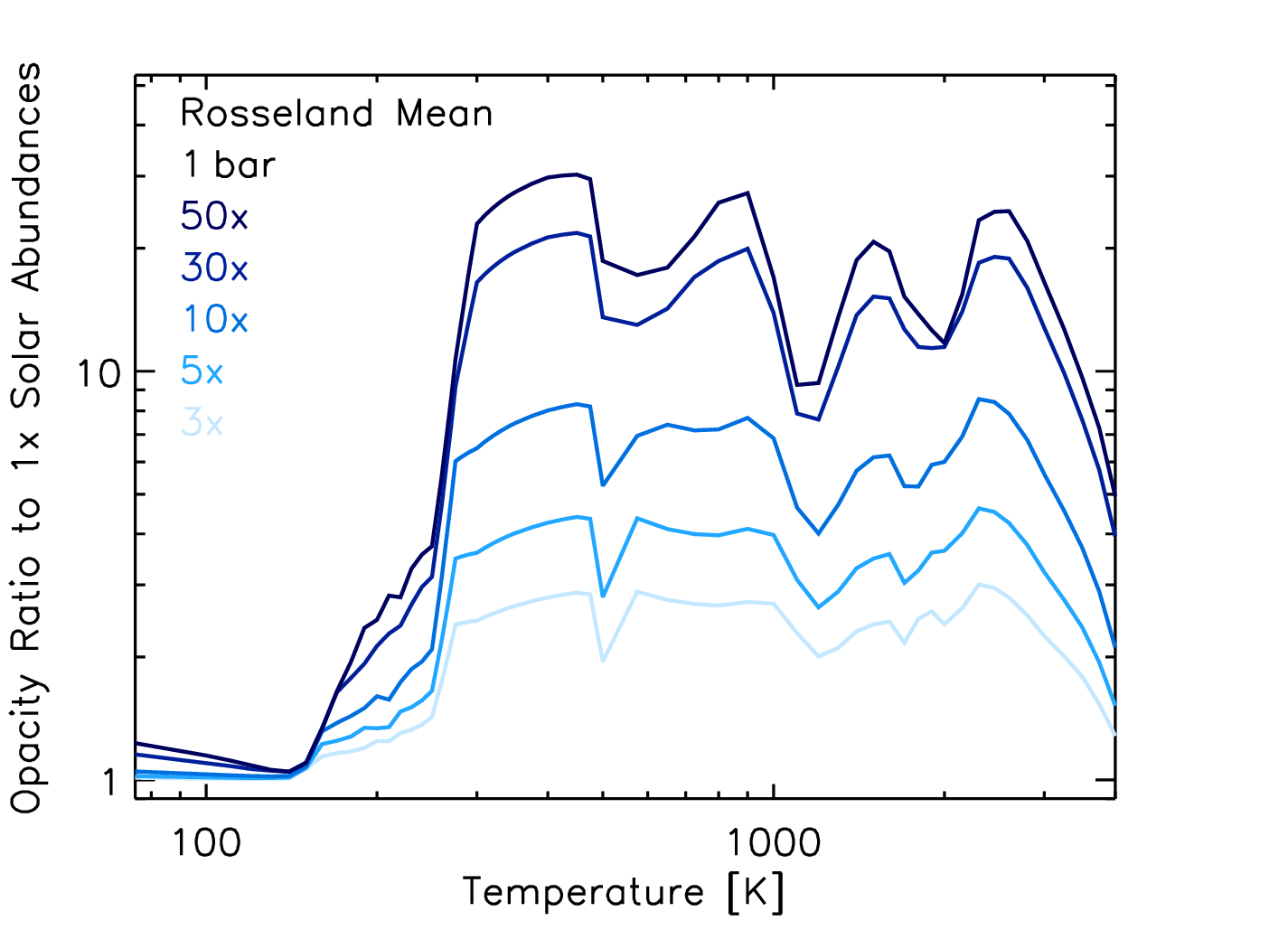}
 \caption{Shown are the local RM opacities at 1 mbar (top) and 1 bar (bottom) for enhanced metallicity atmospheres.  These are plotted as a ratio to the calculation at solar abundances.  Metallicity increases from 3$\times$ (light blue) to 50$\times$ (dark blue).  From 200 to 3500 K, the local RM opacity increases not quite linearly with metallicity.}
   \label{metals}
\end{center}
\end{figure}

We have calculated RM and PM opacities over the full pressure and temperature grid for these enhanced metallicity cases.  The results for the RM opacities are shown in Figure \ref{metals}.  The general trend of increased metallicity is of course increased opacity.  Over the temperature range from 250 K to 3000 K, the RM is a strong function of metallicity, rising not quite linearly.  At very low temperatures, after the condensation of water, the effects of metallicity are much weaker.  First water vapor and then ammonia are lost, essentially leaving only methane and H$_2$ CIA opacity as the main opacity sources at lower temperature.  In particular below 150 K, the peak of the weighting function is at wavelengths beyond 20 $\mu$m, where H$_2$ CIA opacity dominates over CH$_4$, leading to a minimal dependence of the opacity on metallicity.

\subsection{Analytic Opacity Fit}
\begin{figure}
%\hspace{3in}
\includegraphics[scale=0.36,angle=90]{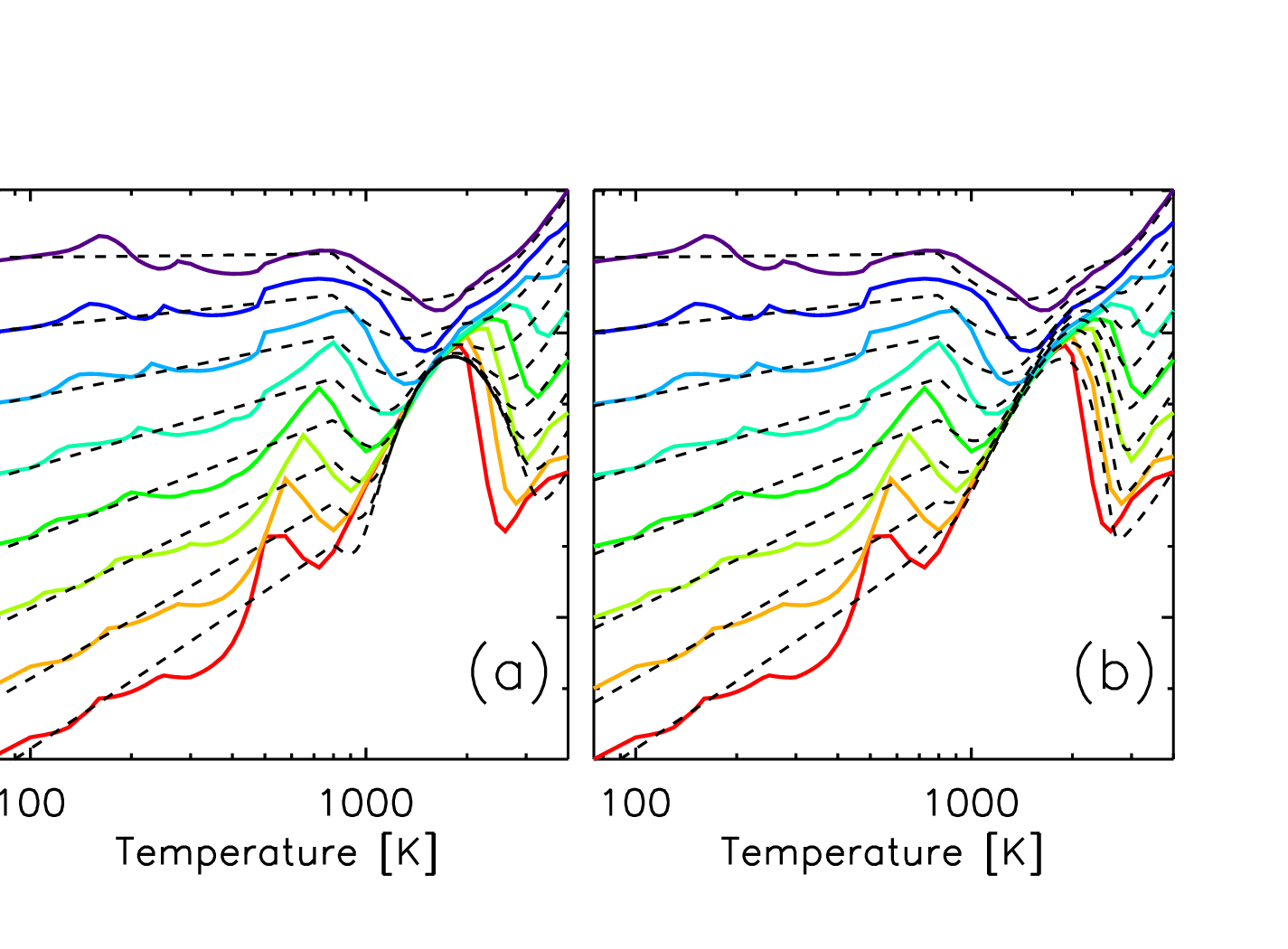}
\caption{RM opacities (solid lines) shown versus local $T$ for a variety of pressures (colors). On the left (a), the analytic fit to the RM opacities described in \ct{Valencia13} is over-plotted as dashed black lines. Similarly, on the right (b) our analytic fit is over-plotted as dashed black lines showing an improvement in the region where $1000\text{ K} \le T \le 3000\text{ K}$. Pressure increases from red to purple.
\label{fit}}
\end{figure}

Recently \ct{Valencia13} used an older version of these tabulated mean opacities at metallicities from solar, 30$\times$ solar ([M/H]=+1.5), and 50$\times$ solar ([M/H]=+1.7) to derive analytic fits to the RM opacities as a function of pressure, temperature, and metallicity.  While interpolation in tables is certainly more accurate, there are instances where the speed of an analytic fit is more important than high accuracy.  The behavior of opacities is quite complex and it is difficult to capture the full behavior with an analytic fit.  Impressively, \ct{Valencia13} were able to provide a reasonable fit to the RM opacities with analytic equations that used 11 coefficients fit by least-squares minimization, by breaking the fit into regions above and below 800 K.  Their fit is shown in Figure \ref{fit}\emph{a}.

Their fit is generally good, but somewhat fails to reproduce the asymmetrical shape of the opacity maximum from 800-3000 K seen at pressures below $\sim$ 1 bar.  We find that we can generally reproduce this feature better by replacing the quadratic term in \ct{Valencia13} (their equation 2) with two terms, one an inverse tangent, and one an exponential with a pressure dependent term.  The cost is relatively minor, with the inclusion of two addition coefficients, making our fit need 13 coefficients, instead of the 11 used in \ct{Valencia13}.

\begin{figure}
%\hspace{3in}
\includegraphics[width=2.70in,angle=90]{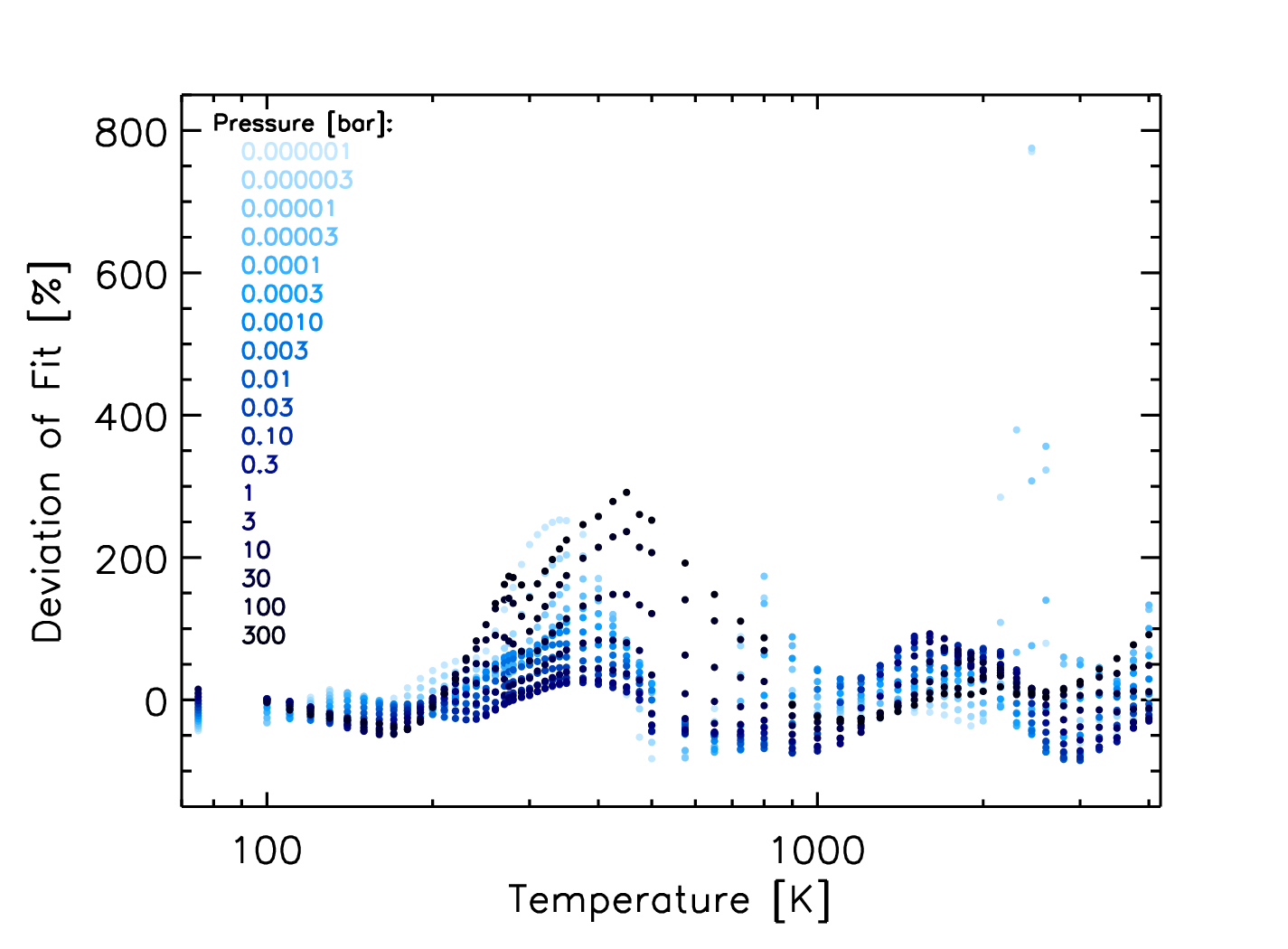}
\caption{This plot shows the percent deviations between the RM opacity tabulation, at solar metallicity, and the analytic fit shown in Figure \ref{fit}b, with coefficients tabulated in Table 2.  The fit reproduces the tables within a factor of 2 for 90\% of the points at solar metallicity, and 80\% at [M/H]=+1.7.
\label{errfit}}
\end{figure}

The analytic RM fit for the opacity $\kappa_{\text{gas}}$ (in cm$^2$ g$^{-1}$) is the sum of two components that have separate analytic fits,  $\kappa_{\text{lowP}}$ and $\kappa_{\text{highP}}$, that are related as:

\begin{equation}
\kappa_{\text{gas}} = \kappa_{\text{lowP}} + \kappa_{\text{highP}}
\end{equation}

where we define,

\begin{eqnarray}
\text{log}_{10}\kappa_{\text{lowP}} = c_1\text{tan}^{-1}\left(\text{log}_{10}T-c_2\right) - \\ \frac{c_3}{\text{log}_{10}P+c_4}e^{\left( \text{log}_{10}T-c_5 \right)^2} +c_6\text{met} + c_7 \nonumber
\label{lowP}
\end{eqnarray}
and
\begin{eqnarray}
\text{log}_{10}\kappa_{\text{highP}} = c_8 + c_9\text{log}_{10}T + \\ c_{10}(\text{log}_{10}T)^2 + \text{log}_{10}P(c_{11}+c_{12}\text{log}_{10}T) + \nonumber \\ c_{13}\text{met}\left[\frac{1}{2} + \frac{1}{\pi} \text{tan}^{-1} \left( \frac{\text{log}_{10}T - 2.5}{0.2} \nonumber \right) \right]
\label{highP}
\end{eqnarray}
with $T$ in K, $P$ in dyne cm$^{-2}$, and ``met'' as the metallicity, [M/H].  The coefficients $c_1$ through $c_{13}$ are given in Table 2.  The fit error at solar metallicity is shown in Figure \ref{errfit}.  Differences are typically on the order of 50\%.  The fit reproduces the tables within a factor of 2 for 90\% of the points at solar metallicity, and 80\% at [M/H]=+1.7.  For some users the flexibility of an analytic fit may be more important than accuracy.

%
%\begin{table}[htdp]
%\caption{Coefficients used for opacity fit}
%\begin{center}
%\begin{tabular}{|cc|ccc|} \hline
% &For all $T$ &  & $T \textless 800$ K & $T \textgreater 800$ K \\ \hline 
%$c_1$ & $10.30549$ & $c_8$ & $-14.051$ & $82.241$ \\
%$c_2$ & $2.88033$ & $c_9$ & $3.055$ & $-55.456$ \\
%$c_3$ & $6.270705\times10^{-15}$ & $c_{10}$ & $0.024$ & $8.754$ \\
%$c_4$ & $3.15623$ & $c_{11}$ & $1.877$ & $0.7048$ \\
%$c_5$ & $-2.519707$ & $c_{12}$ & $-0.445$ & $-0.0414$ \\
%$c_6$ & $0.84315$ & $c_{13}$ & $0.8321$ & $0.8321$ \\
%$c_7$ & $-5.475555$ &  &  &  \\ \hline
%\end{tabular}
%\end{center}
%\label{default}
%\end{table}%
%
\section{Effects of Weighting Temperature} \label{weight}
Over the past several years a variety of analytic work on the temperature structure of strongly irradiated planetary atmospheres has been published \cp[e.g.,][]{Hansen08,Guillot10,Heng12,Robinson12,Parmentier14a,Parmentier14b}.  These works generally aim to understand the atmospheric temperature structure, and presence or absence of temperature inversions, with as few free parameters as possible.  Generally at a minimum one needs to know the thermal infrared opacity, and the relevant ``visible'' opacity for incoming stellar intensity.  The nature of the visible opacity is in principle straightforward to physically understand.  What one is interested in is how the stellar radiation field is attenuated by the wavelength-dependent opacity of the planetary atmosphere, as a function for depth.  Often this is prescribed as a gray atmospheric constant, meaning the visible opacity is chosen as a wavelength-independent constant \cp[e.g.][]{Heng12} or that the ratio of the visible to thermal opacity is a constant \cp[e.g.][]{Guillot10}.  In principle, one could be interested in an opacity based either on the RM or PM, but we caution that either is an approximation to the real atmospheric bevahior, as the wavelength-dependent downwelling (and upwelling) stellar intensity can in general change strongly with depth.

Within the framework of an opacity table theses RM or PM visible opacities can be readily calculated across our \emph{P-T} space by modifying the RM and PM formulas to use the temperature of a \emph{stellar} \teff, rather than the \emph{local} gas temperature.  Hence the weighting function is no longer the intensity characteristic of the local temperature but that of the incoming stellar intensity.  In practice here we use stellar blackbody intensities from 3000 to 7000 K, and the same tabulated wavelength-dependent opacities, including scattering.

These visible opacities should be used with some caution.  \ct{Guillot10} (his equation 9) postulate a use for the PM for the incident stellar intensity at the top of the atmosphere, however, the stellar intensity begins to dramatically deviate from a blackbody as the wavelength dependent flux is carved by absorption, with increasing depth.  However, in a revision of the \ct{Guillot10} work, \ct{Parmentier14b} (see their \S 5.1.1) suggest that instead the stellar-\teff\ weighted RM is much more useful to understand the depth to which incident stellar flux penentrates, since the RM most heavily weights at wavelengths where opacity is low.  They base their suggestion of the utility of the weighted RM by comparing derived model temperature structures that utilize gray opacities, to those that use full frequency-dependent opacity calculations.  For our purposes, we choose to calculate both the stellar-weighted RM and PM tables over our large \emph{P-T} space, such that they can be of some use to the community, for current and future investigations.
 \begin{figure}[ht] 
\begin{center}
 \includegraphics[width=2.70in,angle=90]{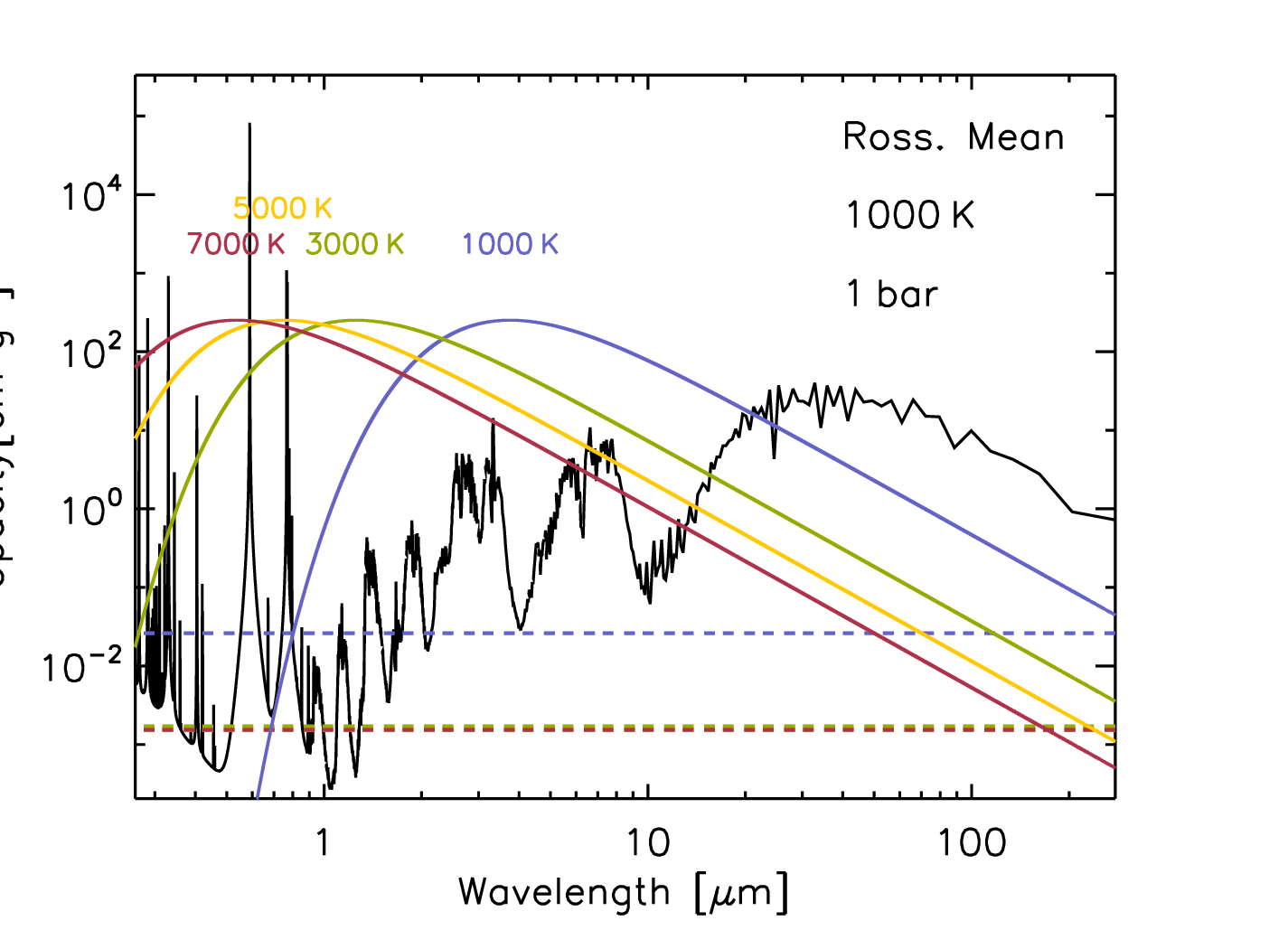}
 \includegraphics[width=2.70in,angle=90]{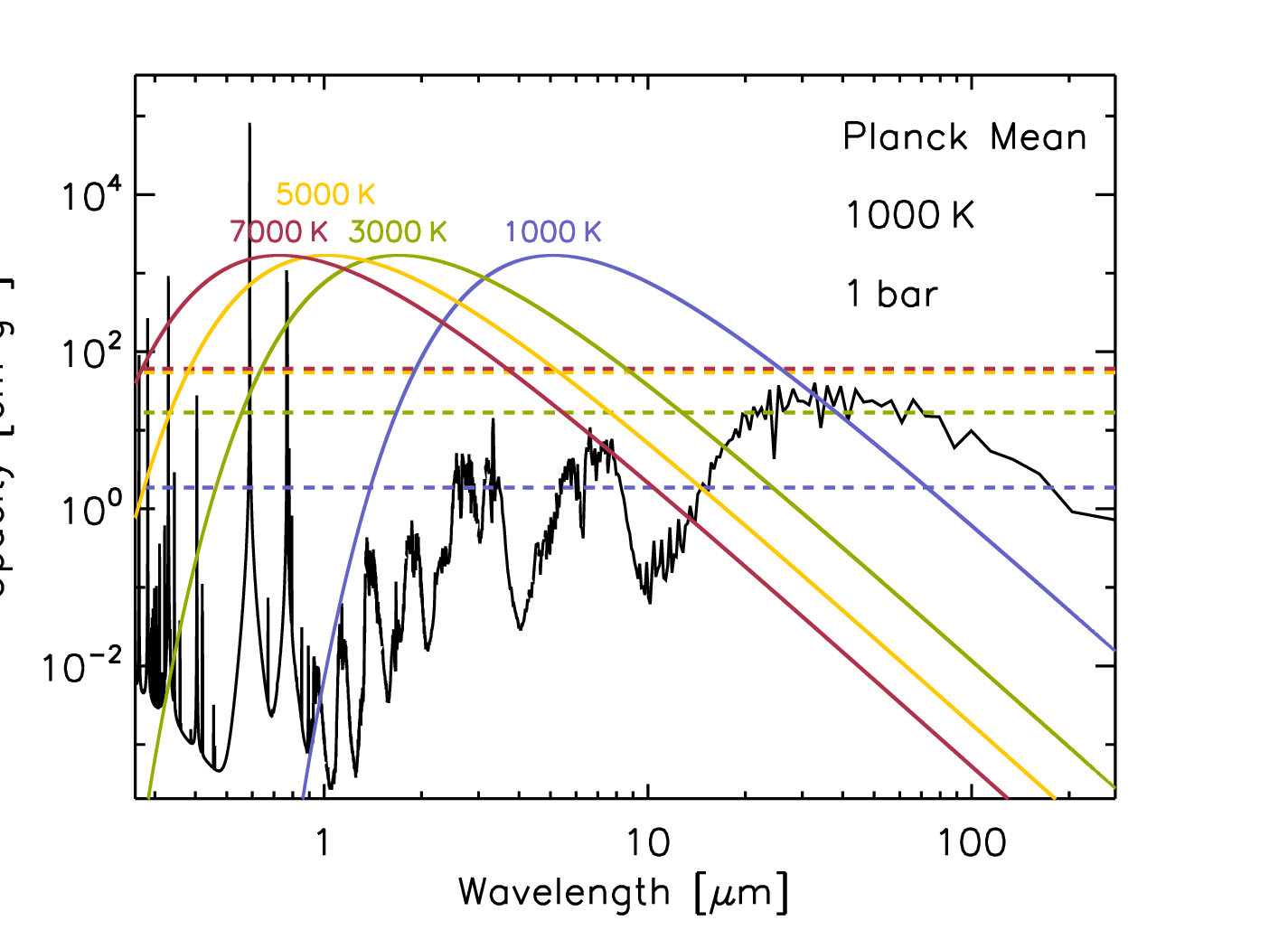}
 \caption{Examples of RM opacity (top) and PM opacity (bottom) at 1000 K and 1 bar.  Shown are the weighting functions at 1000 K (blue), 3000 K (green), 5000 K (yellow), and 7000 K (red).  The RM and PM opacities are shown as the corresponding dashed lines in these same colors.  The wavelength dependent total gaseous opacity is shown in black.  From 3000 to 7000 K the RM is strongly effected by opacity minima at optical wavelengths, while the PM is strongly effected by opacity maxima at optical wavelengths.}
   \label{weight}
\end{center}
\end{figure}

This change in the temperature of the weighting function can lead to dramatically different results for the RM and PM for a given composition.  An example at 1 bar and 1000 K is shown in Figure \ref{weight}.  In the top panel, as the weighting function moves from the near infrared (1000 K) towards the optical (3000 K and higher) the RM falls dramatically due to opacity minima between the pressure-broadened alkali lines, which dominates the RM calculation.  In the bottom panel, for the PM, somewhat opposite behavior is found.  As the weighting function overlaps with these same alkali lines they again dominate, leading to a dramatic increase in the PM opacity.

\begin{figure}[ht] 
\begin{center}
 \includegraphics[width=2.70in,angle=90]{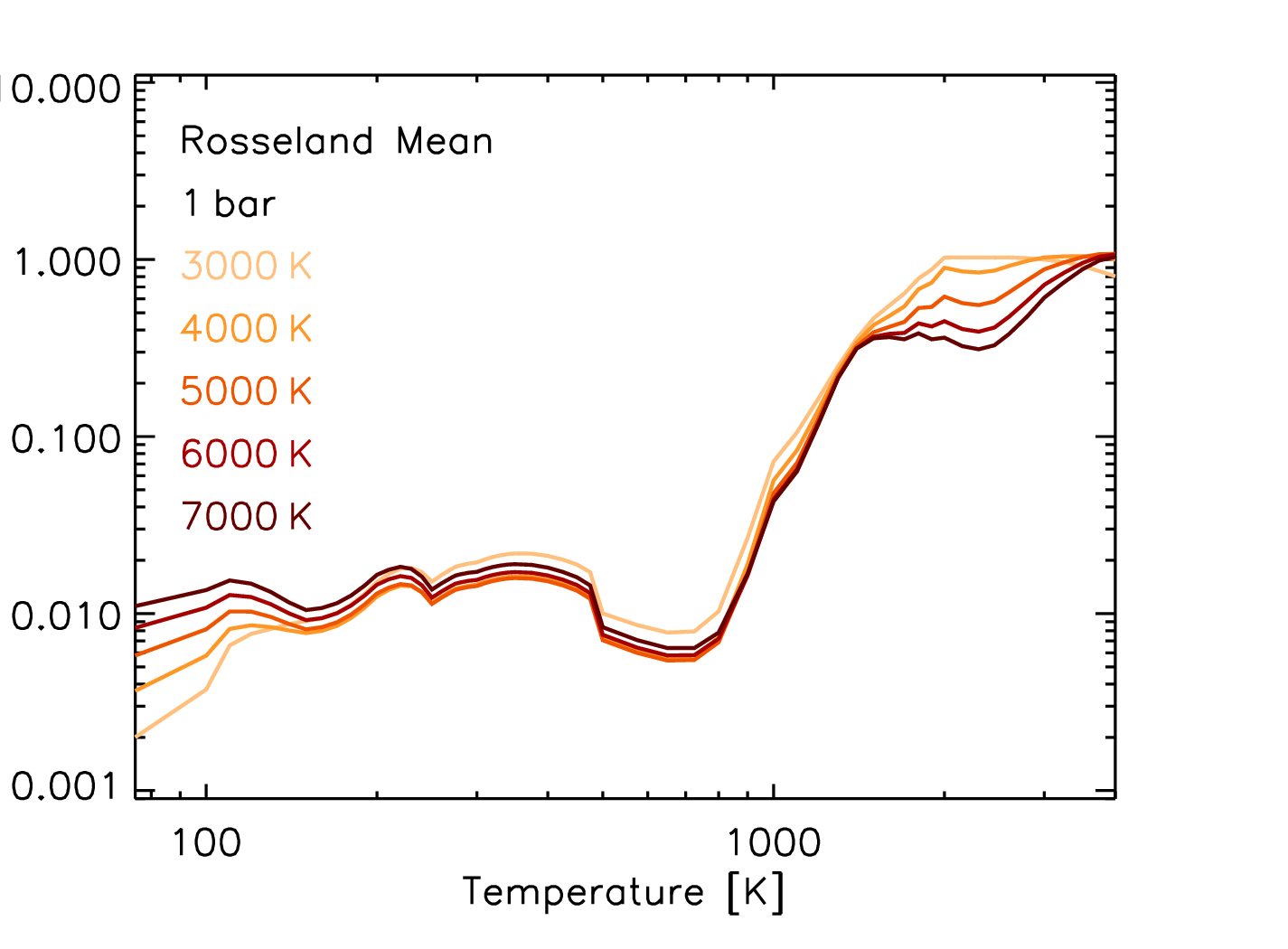}
 \caption{RM opacity ratios at 1 bar.  Calculations are shown for weighting temperatures from 3000 to 7000 K, shown as a ratio to calculations made at the local temperature.}
   \label{weight2}
\end{center}
\end{figure}

Some general behavior is shown in Figure \ref{weight2}.  This is a plot of the ratio of RM calculated at the stellar temperature, divided by the RM calculated at the local temperature, at 1 bar.   A change in behavior is found around 1000 K.  At higher temperatures, there is a much weaker dependence on the weighting function temperature, since the local temperature weighting function moves toward having a greater overlap with the stellar \teff s.  The ratio approaches 1 and is exactly 1.0 for 3000 K, with a 3000 K \teff, and at 4000 K, for a 4000 K \teff, as expected.  Below 1000 K, even when the gas becomes cool enough that Na and K condense and are lost from the gas phase, the optical opacity windows do not change dramatically, and water vapor dominates in the mid IR for the thermal opacity, such that the behavior does not have a strong dependence on the local gas temperature.

\section{Discussion and Conclusions} 
We present our calculated opacities in Tables 3-8, which run from solar metallicity ([M/H]=0.0) to $\sim$50$\times$ solar ([M/H]=+1.7), including intermediate increments of +0.5, +0.7, +1.0, and +1.5.  The tables are over the same grid in temperature $T$ in K, and pressure $P$ in dyne cm$^{-2}$.  The ideal gas law is assumed, and using the appropriate chemical mixing ratios the density $\rho$ in g cm$^{-3}$ is also given.  At the local temperature $T$ the RM and PM opacities are tabulated, in cm$^2$ g$^{-1}$.  Also tabulated are the RM and PM opacities tabulated at stellar incident blackbody weighting temperatures of 3000, 4000, 5000, 6000, and 7000 K (see \S4).

Databases of the opacities at the relatively low temperatures relevant to giant planets and ultracool dwarfs have significantly advanced since the first model atmospheres of these objects nearly 20 years ago.  In our tabulations we use state-of-the art calculations for all molecules and atoms.  The generally good agreement between models of brown dwarf atmospheres and observations \cp{Stephens09,Saumon12,Morley12} suggests that we are on the right track in understanding the chemical abundances and opacity of cool atmospheres.  The general very close agreement at solar metallicity between our calculations here and F08, suggests that future tabulations of purely gaseous opacities over this range of pressure, temperature, and metallicity, probably will not change dramatically in the future.  Future directions could include the use of a variety of C/O ratios, which change the mixing ratios of water vapor, carbon monoxide, and methane.  With regard to the inclusion of cloud opacity in atmospheres, we note that \ct{Cuzzi14} recently published a method to calculate the mean opacities of porous aggregates for use in dusty disks and cloudy atmospheres.
 
With regard to main absorbers that could have an identifiable impact on the mean opacities, and where there is still likely considerable room for improvement, we suggest the alkali metals.  Our treatment of alkali metal opacity, following the work of \ct{BMS}, is still approximate.  Detailed calculations of alkali line shapes, perturbed by collisions with H$_2$ and He are still in progress \cp{Allard03,Allard05}, but they do not yet reach to high enough pressures to be used over all of our phase space.  Also, for methane, detailed comparisons should be made between the databases of \ct{Yurchenko14}, used here, and the recent calculations of \ct{Rey14}.

We hope that these calculations will find broad use in the community.  A particularly interesting area of future study would be giant planet formation by core-nucleated accretion \cp{Perri74,Mizuno80,Stevenson82b}.  The concurrent accretion of solids and gas \cp{Pollack96} may allow for a wide range of metallicity enhancements, and subsequent opacity enhancements, that could change as a function of time and location within the accreting H/He envelope \cp{Podolak88,Mordasini06,Fortney13}, and could plausibly lead to new, and previously unexplored behavior in formation models.

\section{Acknowledgments} 
We thank the anonymous referee for helpful comments, as well as V.~Parmentier, T.~Robinson, and T.~Guillot for enlightening discussions on analytic models of gray atmospheres.

\begin{deluxetable*}{ll} 
\tabletypesize{\scriptsize} 
%\tablewidth{0pt} 
\tablecaption{\label{tab:opac}Molecules used for opacity calculations.}
\tablehead{\colhead{Molecule Name} & \colhead{Opacity Source(s)}}
\startdata 
 CH$_4$ &  \citet{Yurchenko13};\citet{Yurchenko14};\citet{Kark94}; HITRAN'08\tablenotemark{a} isotopes; \tablenotemark{b}  \\
 CO & HITEMP'10\tablenotemark{c}; \citet{Tipping:1976} \\
 CO$_2$ &  \citet{Huang14}; \citet{Huang13}\\
 CrH &  \citet{Burrows:2002}  \\
 FeH &  \citet{Dulick:2003}; \citet{Hargreaves:2010} \\
 H$_2$O & \citet{Partridge:1997}; \citet{Gamache:1998}; HITRAN'08\tablenotemark{a} isotopes  \\
 H$_2$S & \citet{Kissel:2002};\citet{Wattson:1992} - private communication; HITRAN'08\tablenotemark{a} isotopes  \\
 H$_2$ & \citet{Richard:2012} - CIA \\
 NH$_3$ & \citet{Yurchenko:2011}; \citet{Nemtchinov:2004} \\
 PH$_3$ &  \citet{Nikitin:2009}; GEISA\tablenotemark{d}; HITRAN'08\tablenotemark{a} \\
 TiO & \citet{Schwenke:1998};\citet{Allard:2000}\\
 VO &  \citet{Alvarez:1998}
\enddata
\tablenotetext{(a)} {\citet{Rothman:2009};\it http://www.cfa.harvard.edu/hitran/updates.html}
\tablenotetext{(b)} {\it http://icb.u-bourgogne.fr/OMR/SMA/SHTDS/STDS.html}
\tablenotetext{(c)} {\citet{Rothman:2010};\it http://www.cfa.harvard.edu/hitran/HITEMP.html}
\tablenotetext{(d)} {\it http://ether.ipsl.jussieu.fr/etherTypo/?id=950}
\end{deluxetable*}

\begin{deluxetable}{cccccc}
\tabletypesize{\scriptsize}
\tablecolumns{15}
\tablewidth{0pc}
\tablecaption{Coefficients used for opacity fit}
\tablehead{
\colhead{} & \colhead{For all $T$} & \colhead{} & \colhead{} & \colhead{$T < 800$ K} & \colhead{$T > 800$ K}} \\
\startdata
$c_{1}$ & 10.602 &&  $c_{8}$ & -14.051 & 82.241 \\
$c_{2}$ & 2.882 && $c_{9}$ & 3.055 & -55.456 \\ 
$c_{3}$ & 6.09$\times 10^{-15}$ && $c_{10}$ & 0.024 & 8.754 \\   
$c_{4}$ & 2.954 && $c_{11}$ & 1.877 & 0.7048 \\
$c_{5}$ & -2.526 && $c_{12}$ & -0.445 & -0.0414 \\
$c_{6}$ & 0.843 && $c_{13}$ & 0.8321 & 0.8321 \\
$c_{7}$ & -5.490 && \nodata & \nodata & \nodata \\
\enddata 
\end{deluxetable}

\begin{turnpage}

\begin{deluxetable*}{ccccccccccccccc}
\tabletypesize{\scriptsize}
\setlength{\tabcolsep}{0.01in} 
\tablecolumns{15}
\tablewidth{0pc}
\tablecaption{Mean Opacities for $\text{[M/H]}  = 0.0$}
\tablehead{
\colhead{} & \colhead{} & \colhead{}   &  \multicolumn{2}{c}{Local $T$} & \multicolumn{2}{c}{$T_{{\text{eff}}}=3000$ K} &  \multicolumn{2}{c}{$T_{\text{eff}}=4000$ K} &  \multicolumn{2}{c}{$T_{\text{eff}}=5000$ K} &  \multicolumn{2}{c}{$T_{\text{eff}}=6000$ K} &  \multicolumn{2}{c}{$T_{\text{eff}}=7000$ K} \\
%\cline{4-5} \cline{6-7} \cline{8-9} \cline{10-11} \cline{12-13} \cline{14-15} \\
\colhead{$T$} & \colhead{$P$}   & \colhead{$\rho$}  & \colhead{$\kappa_{\text{R}}$} & \colhead{$\kappa_{\text{P}}$} & \colhead{$\kappa_{\text{R}}$} & \colhead{$\kappa_{\text{P}}$} & \colhead{$\kappa_{\text{R}}$} & \colhead{$\kappa_{\text{P}}$} &  \colhead{$\kappa_{\text{R}}$} & \colhead{$\kappa_{\text{P}}$}  & \colhead{$\kappa_{\text{R}}$} & \colhead{$\kappa_{\text{P}}$} & \colhead{$\kappa_{\text{R}}$} & \colhead{$\kappa_{\text{P}}$} \\
\colhead{K} & \colhead{dyne cm$^{-2}$}   & \colhead{g cm$^{-3}$}  & \colhead{cm$^{2}$ g$^{-1}$} & \colhead{cm$^{2}$ g$^{-1}$} & \colhead{cm$^{2}$ g$^{-1}$} & \colhead{cm$^{2}$ g$^{-1}$} &  \colhead{cm$^{2}$ g$^{-1}$} & \colhead{cm$^{2}$ g$^{-1}$} & \colhead{cm$^{2}$ g$^{-1}$} & \colhead{cm$^{2}$ g$^{-1}$} & \colhead{cm$^{2}$ g$^{-1}$} & \colhead{cm$^{2}$ g$^{-1}$} & \colhead{cm$^{2}$ g$^{-1}$} & \colhead{cm$^{2}$ g$^{-1}$}} 
\startdata
  75 & 1E+00 & 3.759E-10 & 1.019E-08 & 2.747E-05 & 8.173E-07 & 1.177E-01 & 1.814E-06 & 6.274E-02 & 3.371E-06 & 3.711E-02 & 5.497E-06 & 2.401E-02 & 8.088E-06 & 1.680E-02 \\
  75 & 3E+00 & 1.127E-09 & 3.037E-08 & 9.568E-06 & 9.905E-07 & 1.177E-01 & 2.178E-06 & 6.273E-02 & 4.027E-06 & 3.710E-02 & 6.542E-06 & 2.401E-02 & 9.597E-06 & 1.680E-02 \\
  75 & 1E+01 & 3.759E-09 & 1.007E-07 & 3.431E-06 & 1.200E-06 & 1.177E-01 & 2.618E-06 & 6.274E-02 & 4.812E-06 & 3.711E-02 & 7.787E-06 & 2.401E-02 & 1.139E-05 & 1.680E-02 \\
  75 & 3E+01 & 1.127E-08 & 3.015E-07 & 2.043E-06 & 1.426E-06 & 1.177E-01 & 3.088E-06 & 6.275E-02 & 5.648E-06 & 3.712E-02 & 9.106E-06 & 2.401E-02 & 1.328E-05 & 1.680E-02 \\
  75 & 1E+02 & 3.759E-08 & 1.003E-06 & 2.833E-06 & 1.751E-06 & 1.179E-01 & 3.761E-06 & 6.281E-02 & 6.839E-06 & 3.715E-02 & 1.097E-05 & 2.404E-02 & 1.595E-05 & 1.682E-02 \\
  75 & 3E+02 & 1.127E-07 & 3.006E-06 & 6.726E-06 & 2.181E-06 & 1.182E-01 & 4.642E-06 & 6.299E-02 & 8.381E-06 & 3.725E-02 & 1.337E-05 & 2.410E-02 & 1.935E-05 & 1.686E-02 \\
  75 & 1E+03 & 3.759E-07 & 1.001E-05 & 2.105E-05 & 2.907E-06 & 1.193E-01 & 6.098E-06 & 6.356E-02 & 1.089E-05 & 3.758E-02 & 1.723E-05 & 2.431E-02 & 2.476E-05 & 1.700E-02 \\
  75 & 3E+03 & 1.127E-06 & 3.002E-05 & 6.239E-05 & 3.950E-06 & 1.222E-01 & 8.136E-06 & 6.501E-02 & 1.432E-05 & 3.841E-02 & 2.242E-05 & 2.483E-02 & 3.194E-05 & 1.\text{eff}E-02 \\
  75 & 1E+04 & 3.759E-06 & 1.000E-04 & 2.068E-04 & 5.729E-06 & 1.285E-01 & 1.148E-05 & 6.824E-02 & 1.981E-05 & 4.026E-02 & 3.052E-05 & 2.599E-02 & 4.294E-05 & 1.815E-02 \\
  75 & 3E+04 & 1.127E-05 & 2.999E-04 & 6.190E-04 & 7.997E-06 & 1.328E-01 & 1.561E-05 & 7.038E-02 & 2.637E-05 & 4.148E-02 & 3.996E-05 & 2.676E-02 & 5.548E-05 & 1.867E-02 \\
  75 & 1E+05 & 3.759E-05 & 9.993E-04 & 2.061E-03 & 1.106E-05 & 1.303E-01 & 2.108E-05 & 6.900E-02 & 3.486E-05 & 4.065E-02 & 5.188E-05 & 2.622E-02 & 7.100E-05 & 1.830E-02 \\
  75 & 3E+05 & 1.127E-04 & 2.997E-03 & 6.182E-03 & 1.448E-05 & 1.300E-01 & 2.709E-05 & 6.884E-02 & 4.399E-05 & 4.057E-02 & 6.442E-05 & 2.618E-02 & 8.701E-05 & 1.828E-02 \\
  75 & 1E+06 & 3.759E-04 & 9.989E-03 & 2.060E-02 & 1.999E-05 & 1.329E-01 & 3.655E-05 & 7.056E-02 & 5.788E-05 & 4.166E-02 & 8.292E-05 & 2.691E-02 & 1.100E-04 & 1.879E-02 \\
  75 & 3E+06 & 1.127E-03 & 2.996E-02 & 6.182E-02 & 3.672E-05 & 1.411E-01 & 6.178E-05 & 7.550E-02 & 9.108E-05 & 4.476E-02 & 1.236E-04 & 2.898E-02 & 1.575E-04 & 2.026E-02 \\
  75 & 1E+07 & 3.759E-03 & 9.988E-02 & 2.060E-01 & 1.119E-04 & 1.701E-01 & 1.401E-04 & 9.277E-02 & 1.712E-04 & 5.563E-02 & 2.068E-04 & 3.626E-02 & 2.447E-04 & 2.542E-02 \\
 100 & 1E+00 & 2.821E-10 & 2.071E-08 & 3.332E-01 & 1.336E-06 & 1.432E-01 & 2.942E-06 & 7.613E-02 & 5.432E-06 & 4.501E-02 & 8.809E-06 & 2.910E-02 & 1.289E-05 & 2.032E-02
\enddata
\tablecomments{Table 3 is published in its entirety in the electronic edition of the \textit{Astrophysical Journal Supplement}. A portion is shown here for guidance regarding its form and content.  $\kappa_{\text{R}}$ is the Rosseland Mean opacity and $\kappa_{\text{P}}$ is the Planck Mean opacity. Columns showing $T_{\rm eff}$ values of 3000-7000 K use this temperature, instead of the local temperature, in the weighting function.  The online version includes an additional significant figure for the density and opacities.}
\end{deluxetable*}

\begin{deluxetable}{ccccccccccccccc}
\tabletypesize{\scriptsize}
\setlength{\tabcolsep}{0.01in} 
\tablecolumns{15}
\tablewidth{0pc}
\tablecaption{Mean Opacities for $\text{[M/H]}  = 0.5$}
\tablehead{
\colhead{} & \colhead{} & \colhead{}   &  \multicolumn{2}{c}{Local $T$} & \multicolumn{2}{c}{$T_{{\text{eff}}}=3000$ K} &  \multicolumn{2}{c}{$T_{\text{eff}}=4000$ K} &  \multicolumn{2}{c}{$T_{\text{eff}}=5000$ K} &  \multicolumn{2}{c}{$T_{\text{eff}}=6000$ K} &  \multicolumn{2}{c}{$T_{\text{eff}}=7000$ K} \\
%\cline{4-5} \cline{6-7} \cline{8-9} \cline{10-11} \cline{12-13} \cline{14-15} \\
\colhead{$T$} & \colhead{$P$}   & \colhead{$\rho$}  & \colhead{$\kappa_{\text{R}}$} & \colhead{$\kappa_{\text{P}}$} & \colhead{$\kappa_{\text{R}}$} & \colhead{$\kappa_{\text{P}}$} & \colhead{$\kappa_{\text{R}}$} & \colhead{$\kappa_{\text{P}}$} &  \colhead{$\kappa_{\text{R}}$} & \colhead{$\kappa_{\text{P}}$}  & \colhead{$\kappa_{\text{R}}$} & \colhead{$\kappa_{\text{P}}$} & \colhead{$\kappa_{\text{R}}$} & \colhead{$\kappa_{\text{P}}$} \\
\colhead{K} & \colhead{dyne cm$^{-2}$}   & \colhead{g cm$^{-3}$}  & \colhead{cm$^{2}$ g$^{-1}$} & \colhead{cm$^{2}$ g$^{-1}$} & \colhead{cm$^{2}$ g$^{-1}$} & \colhead{cm$^{2}$ g$^{-1}$} &  \colhead{cm$^{2}$ g$^{-1}$} & \colhead{cm$^{2}$ g$^{-1}$} & \colhead{cm$^{2}$ g$^{-1}$} & \colhead{cm$^{2}$ g$^{-1}$} & \colhead{cm$^{2}$ g$^{-1}$} & \colhead{cm$^{2}$ g$^{-1}$} & \colhead{cm$^{2}$ g$^{-1}$} & \colhead{cm$^{2}$ g$^{-1}$}} 
\startdata
  75 & 1E+00 & 3.784E-10 & 1.033E-08 & 2.777E-05 & 8.651E-07 & 3.708E-01 & 1.924E-06 & 1.974E-01 & 3.582E-06 & 1.164E-01 & 5.847E-06 & 7.490E-02 & 8.609E-06 & 5.186E-02 \\
  75 & 3E+00 & 1.135E-09 & 3.076E-08 & 9.997E-06 & 1.066E-06 & 3.708E-01 & 2.351E-06 & 1.974E-01 & 4.351E-06 & 1.164E-01 & 7.076E-06 & 7.491E-02 & 1.038E-05 & 5.186E-02 \\
  75 & 1E+01 & 3.784E-09 & 1.020E-07 & 3.910E-06 & 1.304E-06 & 3.709E-01 & 2.851E-06 & 1.974E-01 & 5.248E-06 & 1.164E-01 & 8.500E-06 & 7.491E-02 & 1.244E-05 & 5.187E-02 \\
  75 & 3E+01 & 1.135E-08 & 3.054E-07 & 2.549E-06 & 1.537E-06 & 3.710E-01 & 3.341E-06 & 1.975E-01 & 6.125E-06 & 1.165E-01 & 9.888E-06 & 7.493E-02 & 1.443E-05 & 5.188E-02 \\
  75 & 1E+02 & 3.784E-08 & 1.016E-06 & 3.394E-06 & 1.861E-06 & 3.713E-01 & 4.020E-06 & 1.977E-01 & 7.334E-06 & 1.166E-01 & 1.179E-05 & 7.500E-02 & 1.716E-05 & 5.193E-02 \\
  75 & 3E+02 & 1.135E-07 & 3.045E-06 & 7.410E-06 & 2.302E-06 & 3.724E-01 & 4.935E-06 & 1.982E-01 & 8.950E-06 & 1.169E-01 & 1.432E-05 & 7.520E-02 & 2.075E-05 & 5.206E-02 \\
  75 & 1E+03 & 3.784E-07 & 1.014E-05 & 2.207E-05 & 3.068E-06 & 3.760E-01 & 6.496E-06 & 2.000E-01 & 1.166E-05 & 1.179E-01 & 1.852E-05 & 7.585E-02 & 2.667E-05 & 5.251E-02 \\
  75 & 3E+03 & 1.135E-06 & 3.041E-05 & 6.425E-05 & 4.193E-06 & 3.850E-01 & 8.732E-06 & 2.046E-01 & 1.548E-05 & 1.206E-01 & 2.433E-05 & 7.751E-02 & 3.473E-05 & 5.363E-02 \\
  75 & 1E+04 & 3.784E-06 & 1.013E-04 & 2.108E-04 & 6.142E-06 & 4.049E-01 & 1.248E-05 & 2.147E-01 & 2.171E-05 & 1.264E-01 & 3.359E-05 & 8.116E-02 & 4.735E-05 & 5.612E-02 \\
  75 & 3E+04 & 1.135E-05 & 3.037E-04 & 6.276E-04 & 8.640E-06 & 4.182E-01 & 1.715E-05 & 2.214E-01 & 2.925E-05 & 1.301E-01 & 4.451E-05 & 8.353E-02 & 6.190E-05 & 5.772E-02 \\
  75 & 1E+05 & 3.784E-05 & 1.011E-03 & 2.085E-03 & 1.201E-05 & 4.099E-01 & 2.334E-05 & 2.166E-01 & 3.900E-05 & 1.273E-01 & 5.830E-05 & 8.168E-02 & 7.987E-05 & 5.644E-02 \\
  75 & 3E+05 & 1.135E-04 & 3.034E-03 & 6.252E-03 & 1.574E-05 & 4.069E-01 & 3.010E-05 & 2.151E-01 & 4.939E-05 & 1.264E-01 & 7.260E-05 & 8.111E-02 & 9.809E-05 & 5.605E-02 \\
  75 & 1E+06 & 3.784E-04 & 1.011E-02 & 2.083E-02 & 2.177E-05 & 4.097E-01 & 4.071E-05 & 2.168E-01 & 6.509E-05 & 1.274E-01 & 9.346E-05 & 8.182E-02 & 1.238E-04 & 5.655E-02 \\
  75 & 3E+06 & 1.135E-03 & 3.032E-02 & 6.250E-02 & 4.226E-05 & 4.179E-01 & 7.231E-05 & 2.217E-01 & 1.065E-04 & 1.305E-01 & 1.435E-04 & 8.388E-02 & 1.815E-04 & 5.801E-02 \\
  75 & 1E+07 & 3.784E-03 & 1.010E-01 & 2.083E-01 & 1.508E-04 & 4.466E-01 & 1.807E-04 & 2.388E-01 & 2.116E-04 & 1.413E-01 & 2.480E-04 & 9.108E-02 & 2.871E-04 & 6.312E-02 \\
100 & 1E+00 & 2.844E-10 & 2.348E-08 & 9.863E-01 & 1.567E-06 & 4.472E-01 & 3.455E-06 & 2.375E-01 & 6.384E-06 & 1.401E-01 & 1.035E-05 & 9.014E-02 & 1.515E-05 & 6.240E-02
\enddata
\tablecomments{Table 4 is published in its entirety in the electronic edition of the \textit{Astrophysical Journal Supplement}. A portion is shown here for guidance regarding its form and content.  $\kappa_{\text{R}}$ is the Rosseland Mean opacity and $\kappa_{\text{P}}$ is the Planck Mean opacity. Columns showing $T_{\rm eff}$ values of 3000-7000 K use this temperature, instead of the local temperature, in the weighting function.  The online version includes an additional significant figure for the density and opacities.}
\end{deluxetable}

\begin{deluxetable}{ccccccccccccccc}
\tabletypesize{\scriptsize}
\setlength{\tabcolsep}{0.01in} 
\tablecolumns{15}
\tablewidth{0pc}
\tablecaption{Mean Opacities for $\text{[M/H]}  = 0.7$}
\tablehead{
\colhead{} & \colhead{} & \colhead{}   &  \multicolumn{2}{c}{Local $T$} & \multicolumn{2}{c}{$T_{{\text{eff}}}=3000$ K} &  \multicolumn{2}{c}{$T_{\text{eff}}=4000$ K} &  \multicolumn{2}{c}{$T_{\text{eff}}=5000$ K} &  \multicolumn{2}{c}{$T_{\text{eff}}=6000$ K} &  \multicolumn{2}{c}{$T_{\text{eff}}=7000$ K} \\
%\cline{4-5} \cline{6-7} \cline{8-9} \cline{10-11} \cline{12-13} \cline{14-15} \\
\colhead{$T$} & \colhead{$P$}   & \colhead{$\rho$}  & \colhead{$\kappa_{\text{R}}$} & \colhead{$\kappa_{\text{P}}$} & \colhead{$\kappa_{\text{R}}$} & \colhead{$\kappa_{\text{P}}$} & \colhead{$\kappa_{\text{R}}$} & \colhead{$\kappa_{\text{P}}$} &  \colhead{$\kappa_{\text{R}}$} & \colhead{$\kappa_{\text{P}}$}  & \colhead{$\kappa_{\text{R}}$} & \colhead{$\kappa_{\text{P}}$} & \colhead{$\kappa_{\text{R}}$} & \colhead{$\kappa_{\text{P}}$} \\
\colhead{K} & \colhead{dyne cm$^{-2}$}   & \colhead{g cm$^{-3}$}  & \colhead{cm$^{2}$ g$^{-1}$} & \colhead{cm$^{2}$ g$^{-1}$} & \colhead{cm$^{2}$ g$^{-1}$} & \colhead{cm$^{2}$ g$^{-1}$} &  \colhead{cm$^{2}$ g$^{-1}$} & \colhead{cm$^{2}$ g$^{-1}$} & \colhead{cm$^{2}$ g$^{-1}$} & \colhead{cm$^{2}$ g$^{-1}$} & \colhead{cm$^{2}$ g$^{-1}$} & \colhead{cm$^{2}$ g$^{-1}$} & \colhead{cm$^{2}$ g$^{-1}$} & \colhead{cm$^{2}$ g$^{-1}$}} 
\startdata
  75 & 1E+00 & 3.806E-10 & 1.043E-08 & 2.816E-05 & 8.848E-07 & 5.859E-01 & 1.970E-06 & 3.119E-01 & 3.669E-06 & 1.839E-01 & 5.993E-06 & 1.181E-01 & 8.825E-06 & 8.166E-02 \\
  75 & 3E+00 & 1.142E-09 & 3.108E-08 & 1.036E-05 & 1.098E-06 & 5.860E-01 & 2.423E-06 & 3.119E-01 & 4.487E-06 & 1.839E-01 & 7.299E-06 & 1.181E-01 & 1.071E-05 & 8.167E-02 \\
  75 & 1E+01 & 3.806E-09 & 1.031E-07 & 4.317E-06 & 1.344E-06 & 5.860E-01 & 2.941E-06 & 3.119E-01 & 5.418E-06 & 1.839E-01 & 8.780E-06 & 1.181E-01 & 1.285E-05 & 8.167E-02 \\
  75 & 3E+01 & 1.142E-08 & 3.086E-07 & 2.979E-06 & 1.575E-06 & 5.862E-01 & 3.429E-06 & 3.120E-01 & 6.294E-06 & 1.840E-01 & 1.016E-05 & 1.182E-01 & 1.484E-05 & 8.169E-02 \\
  75 & 1E+02 & 3.806E-08 & 1.026E-06 & 3.875E-06 & 1.895E-06 & 5.854E-01 & 4.103E-06 & 3.116E-01 & 7.499E-06 & 1.837E-01 & 1.207E-05 & 1.180E-01 & 1.757E-05 & 8.157E-02 \\
  75 & 3E+02 & 1.142E-07 & 3.077E-06 & 7.990E-06 & 2.339E-06 & 5.870E-01 & 5.029E-06 & 3.124E-01 & 9.140E-06 & 1.842E-01 & 1.464E-05 & 1.183E-01 & 2.124E-05 & 8.178E-02 \\
  75 & 1E+03 & 3.806E-07 & 1.024E-05 & 2.293E-05 & 3.118E-06 & 5.927E-01 & 6.628E-06 & 3.153E-01 & 1.193E-05 & 1.858E-01 & 1.898E-05 & 1.193E-01 & 2.\text{eff}E-05 & 8.249E-02 \\
  75 & 3E+03 & 1.142E-06 & 3.072E-05 & 6.582E-05 & 4.275E-06 & 6.069E-01 & 8.944E-06 & 3.225E-01 & 1.590E-05 & 1.900E-01 & 2.504E-05 & 1.219E-01 & 3.578E-05 & 8.426E-02 \\
  75 & 1E+04 & 3.806E-06 & 1.023E-04 & 2.141E-04 & 6.290E-06 & 6.383E-01 & 1.286E-05 & 3.385E-01 & 2.245E-05 & 1.991E-01 & 3.480E-05 & 1.277E-01 & 4.911E-05 & 8.818E-02 \\
  75 & 3E+04 & 1.142E-05 & 3.068E-04 & 6.348E-04 & 8.873E-06 & 6.592E-01 & 1.774E-05 & 3.489E-01 & 3.039E-05 & 2.050E-01 & 4.635E-05 & 1.314E-01 & 6.453E-05 & 9.069E-02 \\
  75 & 1E+05 & 3.806E-05 & 1.022E-03 & 2.106E-03 & 1.234E-05 & 6.459E-01 & 2.420E-05 & 3.413E-01 & 4.064E-05 & 2.005E-01 & 6.089E-05 & 1.285E-01 & 8.349E-05 & 8.864E-02 \\
  75 & 3E+05 & 1.142E-04 & 3.065E-03 & 6.311E-03 & 1.617E-05 & 6.407E-01 & 3.122E-05 & 3.386E-01 & 5.150E-05 & 1.989E-01 & 7.587E-05 & 1.274E-01 & 1.025E-04 & 8.794E-02 \\
  75 & 1E+06 & 3.806E-04 & 1.021E-02 & 2.102E-02 & 2.239E-05 & 6.435E-01 & 4.228E-05 & 3.403E-01 & 6.793E-05 & 1.999E-01 & 9.770E-05 & 1.281E-01 & 1.294E-04 & 8.844E-02 \\
  75 & 3E+06 & 1.142E-03 & 3.062E-02 & 6.307E-02 & 4.444E-05 & 6.516E-01 & 7.671E-05 & 3.451E-01 & 1.131E-04 & 2.030E-01 & 1.522E-04 & 1.302E-01 & 1.919E-04 & 8.989E-02 \\
  75 & 1E+07 & 3.806E-03 & 1.020E-01 & 2.102E-01 & 1.692E-04 & 6.801E-01 & 1.996E-04 & 3.621E-01 & 2.300E-04 & 2.136E-01 & 2.662E-04 & 1.373E-01 & 3.056E-04 & 9.495E-02 \\
100 & 1E+00 & 2.872E-10 & 2.684E-08 & 3.021E+00 & 1.766E-06 & 7.849E-01 & 3.894E-06 & 4.173E-01 & 7.194E-06 & 2.464E-01 & 1.166E-05 & 1.586E-01 & 1.705E-05 & 1.097E-01
\enddata
\tablecomments{Table 5 is published in its entirety in the electronic edition of the \textit{Astrophysical Journal Supplement}. A portion is shown here for guidance regarding its form and content.  $\kappa_{\text{R}}$ is the Rosseland Mean opacity and $\kappa_{\text{P}}$ is the Planck Mean opacity. Columns showing $T_{\rm eff}$ values of 3000-7000 K use this temperature, instead of the local temperature, in the weighting function.  The online version includes an additional significant figure for the density and opacities.}
\end{deluxetable}

\begin{deluxetable}{ccccccccccccccc}
\tabletypesize{\scriptsize}
\setlength{\tabcolsep}{0.01in} 
\tablecolumns{15}
\tablewidth{0pc}
\tablecaption{Mean Opacities for $\text{[M/H]}  = 1.0$}
\tablehead{
\colhead{} & \colhead{} & \colhead{}   &  \multicolumn{2}{c}{Local $T$} & \multicolumn{2}{c}{$T_{{\text{eff}}}=3000$ K} &  \multicolumn{2}{c}{$T_{\text{eff}}=4000$ K} &  \multicolumn{2}{c}{$T_{\text{eff}}=5000$ K} &  \multicolumn{2}{c}{$T_{\text{eff}}=6000$ K} &  \multicolumn{2}{c}{$T_{\text{eff}}=7000$ K} \\
%\cline{4-5} \cline{6-7} \cline{8-9} \cline{10-11} \cline{12-13} \cline{14-15} \\
\colhead{$T$} & \colhead{$P$}   & \colhead{$\rho$}  & \colhead{$\kappa_{\text{R}}$} & \colhead{$\kappa_{\text{P}}$} & \colhead{$\kappa_{\text{R}}$} & \colhead{$\kappa_{\text{P}}$} & \colhead{$\kappa_{\text{R}}$} & \colhead{$\kappa_{\text{P}}$} &  \colhead{$\kappa_{\text{R}}$} & \colhead{$\kappa_{\text{P}}$}  & \colhead{$\kappa_{\text{R}}$} & \colhead{$\kappa_{\text{P}}$} & \colhead{$\kappa_{\text{R}}$} & \colhead{$\kappa_{\text{P}}$} \\
\colhead{K} & \colhead{dyne cm$^{-2}$}   & \colhead{g cm$^{-3}$}  & \colhead{cm$^{2}$ g$^{-1}$} & \colhead{cm$^{2}$ g$^{-1}$} & \colhead{cm$^{2}$ g$^{-1}$} & \colhead{cm$^{2}$ g$^{-1}$} &  \colhead{cm$^{2}$ g$^{-1}$} & \colhead{cm$^{2}$ g$^{-1}$} & \colhead{cm$^{2}$ g$^{-1}$} & \colhead{cm$^{2}$ g$^{-1}$} & \colhead{cm$^{2}$ g$^{-1}$} & \colhead{cm$^{2}$ g$^{-1}$} & \colhead{cm$^{2}$ g$^{-1}$} & \colhead{cm$^{2}$ g$^{-1}$}} 
\startdata
  75 & 1E+00 & 3.866E-10 & 1.071E-08 & 2.895E-05 & 9.110E-07 & 1.159E+00 & 2.031E-06 & 6.173E-01 & 3.787E-06 & 3.638E-01 & 6.188E-06 & 2.335E-01 & 9.117E-06 & 1.611E-01 \\
  75 & 3E+00 & 1.160E-09 & 3.191E-08 & 1.134E-05 & 1.140E-06 & 1.159E+00 & 2.518E-06 & 6.173E-01 & 4.669E-06 & 3.638E-01 & 7.600E-06 & 2.335E-01 & 1.116E-05 & 1.611E-01 \\
  75 & 1E+01 & 3.866E-09 & 1.058E-07 & 5.401E-06 & 1.391E-06 & 1.160E+00 & 3.049E-06 & 6.174E-01 & 5.625E-06 & 3.639E-01 & 9.122E-06 & 2.336E-01 & 1.336E-05 & 1.611E-01 \\
  75 & 3E+01 & 1.160E-08 & 3.168E-07 & 4.125E-06 & 1.614E-06 & 1.160E+00 & 3.523E-06 & 6.176E-01 & 6.478E-06 & 3.640E-01 & 1.048E-05 & 2.336E-01 & 1.531E-05 & 1.612E-01 \\
  75 & 1E+02 & 3.866E-08 & 1.054E-06 & 5.147E-06 & 1.925E-06 & 1.161E+00 & 4.184E-06 & 6.182E-01 & 7.667E-06 & 3.643E-01 & 1.236E-05 & 2.338E-01 & 1.802E-05 & 1.613E-01 \\
  75 & 3E+02 & 1.160E-07 & 3.159E-06 & 9.539E-06 & 2.369E-06 & 1.164E+00 & 5.118E-06 & 6.198E-01 & 9.333E-06 & 3.653E-01 & 1.499E-05 & 2.344E-01 & 2.177E-05 & 1.617E-01 \\
  75 & 1E+03 & 3.866E-07 & 1.052E-05 & 2.522E-05 & 3.163E-06 & 1.176E+00 & 6.765E-06 & 6.255E-01 & 1.223E-05 & 3.685E-01 & 1.950E-05 & 2.365E-01 & 2.816E-05 & 1.631E-01 \\
  75 & 3E+03 & 1.160E-06 & 3.155E-05 & 7.002E-05 & 4.358E-06 & 1.202E+00 & 9.187E-06 & 6.390E-01 & 1.642E-05 & 3.763E-01 & 2.593E-05 & 2.413E-01 & 3.712E-05 & 1.664E-01 \\
  75 & 1E+04 & 3.866E-06 & 1.051E-04 & 2.231E-04 & 6.458E-06 & 1.266E+00 & 1.332E-05 & 6.714E-01 & 2.341E-05 & 3.948E-01 & 3.642E-05 & 2.530E-01 & 5.149E-05 & 1.744E-01 \\
  75 & 3E+04 & 1.159E-05 & 3.150E-04 & 6.539E-04 & 9.143E-06 & 1.300E+00 & 1.849E-05 & 6.882E-01 & 3.191E-05 & 4.043E-01 & 4.885E-05 & 2.590E-01 & 6.813E-05 & 1.784E-01 \\
  75 & 1E+05 & 3.866E-05 & 1.049E-03 & 2.161E-03 & 1.272E-05 & 1.281E+00 & 2.529E-05 & 6.770E-01 & 4.282E-05 & 3.975E-01 & 6.442E-05 & 2.545E-01 & 8.849E-05 & 1.753E-01 \\
  75 & 3E+05 & 1.159E-04 & 3.150E-03 & 6.482E-03 & 1.667E-05 & 1.276E+00 & 3.268E-05 & 6.744E-01 & 5.439E-05 & 3.959E-01 & 8.047E-05 & 2.535E-01 & 1.089E-04 & 1.746E-01 \\
  75 & 1E+06 & 3.866E-04 & 1.048E-02 & 2.154E-02 & 2.307E-05 & 1.272E+00 & 4.427E-05 & 6.726E-01 & 7.178E-05 & 3.950E-01 & 1.036E-04 & 2.529E-01 & 1.374E-04 & 1.742E-01 \\
  75 & 3E+06 & 1.159E-03 & 3.142E-02 & 6.460E-02 & 4.745E-05 & 1.277E+00 & 8.320E-05 & 6.757E-01 & 1.232E-04 & 3.970E-01 & 1.654E-04 & 2.543E-01 & 2.078E-04 & 1.752E-01 \\
  75 & 1E+07 & 3.866E-03 & 1.047E-01 & 2.153E-01 & 1.999E-04 & 1.302E+00 & 2.313E-04 & 6.907E-01 & 2.602E-04 & 4.065E-01 & 2.957E-04 & 2.607E-01 & 3.349E-04 & 1.797E-01 \\
  100 & 1E+00 & 2.918E-10 & 2.714E-08 & 3.008E+00 & 1.835E-06 & 1.387E+00 & 4.054E-06 & 7.370E-01 & 7.499E-06 & 4.344E-01 & 1.216E-05 & 2.790E-01 & 1.779E-05 & 1.926E-01
\enddata
\tablecomments{Table 6 is published in its entirety in the electronic edition of the \textit{Astrophysical Journal Supplement}. A portion is shown here for guidance regarding its form and content.  $\kappa_{\text{R}}$ is the Rosseland Mean opacity and $\kappa_{\text{P}}$ is the Planck Mean opacity. Columns showing $T_{\rm eff}$ values of 3000-7000 K use this temperature, instead of the local temperature, in the weighting function.  The online version includes an additional significant figure for the density and opacities.}
\end{deluxetable}

\begin{deluxetable}{ccccccccccccccc}
\tabletypesize{\scriptsize}
\setlength{\tabcolsep}{0.01in} 
\tablecolumns{15}
\tablewidth{0pc}
\tablecaption{Mean Opacities for $\text{[M/H]}  = 1.5$}
\tablehead{
\colhead{} & \colhead{} & \colhead{}   &  \multicolumn{2}{c}{Local $T$} & \multicolumn{2}{c}{$T_{{\text{eff}}}=3000$ K} &  \multicolumn{2}{c}{$T_{\text{eff}}=4000$ K} &  \multicolumn{2}{c}{$T_{\text{eff}}=5000$ K} &  \multicolumn{2}{c}{$T_{\text{eff}}=6000$ K} &  \multicolumn{2}{c}{$T_{\text{eff}}=7000$ K} \\
%\cline{4-5} \cline{6-7} \cline{8-9} \cline{10-11} \cline{12-13} \cline{14-15} \\
\colhead{$T$} & \colhead{$P$}   & \colhead{$\rho$}  & \colhead{$\kappa_{\text{R}}$} & \colhead{$\kappa_{\text{P}}$} & \colhead{$\kappa_{\text{R}}$} & \colhead{$\kappa_{\text{P}}$} & \colhead{$\kappa_{\text{R}}$} & \colhead{$\kappa_{\text{P}}$} &  \colhead{$\kappa_{\text{R}}$} & \colhead{$\kappa_{\text{P}}$}  & \colhead{$\kappa_{\text{R}}$} & \colhead{$\kappa_{\text{P}}$} & \colhead{$\kappa_{\text{R}}$} & \colhead{$\kappa_{\text{P}}$} \\
\colhead{K} & \colhead{dyne cm$^{-2}$}   & \colhead{g cm$^{-3}$}  & \colhead{cm$^{2}$ g$^{-1}$} & \colhead{cm$^{2}$ g$^{-1}$} & \colhead{cm$^{2}$ g$^{-1}$} & \colhead{cm$^{2}$ g$^{-1}$} &  \colhead{cm$^{2}$ g$^{-1}$} & \colhead{cm$^{2}$ g$^{-1}$} & \colhead{cm$^{2}$ g$^{-1}$} & \colhead{cm$^{2}$ g$^{-1}$} & \colhead{cm$^{2}$ g$^{-1}$} & \colhead{cm$^{2}$ g$^{-1}$} & \colhead{cm$^{2}$ g$^{-1}$} & \colhead{cm$^{2}$ g$^{-1}$}} 
\startdata
  75 & 1E+00 & 4.136E-10 & 1.178E-08 & 3.171E-05 & 9.155E-07 & 3.541E+00 & 2.045E-06 & 1.884E+00 & 3.819E-06 & 1.110E+00 & 6.249E-06 & 7.125E-01 & 9.213E-06 & 4.911E-01 \\
  75 & 3E+00 & 1.241E-09 & 3.512E-08 & 1.541E-05 & 1.149E-06 & 3.543E+00 & 2.545E-06 & 1.886E+00 & 4.727E-06 & 1.111E+00 & 7.705E-06 & 7.129E-01 & 1.132E-05 & 4.913E-01 \\
  75 & 1E+01 & 4.137E-09 & 1.165E-07 & 9.907E-06 & 1.381E-06 & 3.544E+00 & 3.039E-06 & 1.886E+00 & 5.622E-06 & 1.111E+00 & 9.136E-06 & 7.129E-01 & 1.339E-05 & 4.914E-01 \\
  75 & 3E+01 & 1.241E-08 & 3.488E-07 & 8.886E-06 & 1.575E-06 & 3.545E+00 & 3.458E-06 & 1.886E+00 & 6.383E-06 & 1.111E+00 & 1.035E-05 & 7.131E-01 & 1.515E-05 & 4.915E-01 \\
  75 & 1E+02 & 4.137E-08 & 1.160E-06 & 1.041E-05 & 1.858E-06 & 3.548E+00 & 4.067E-06 & 1.888E+00 & 7.490E-06 & 1.112E+00 & 1.212E-05 & 7.138E-01 & 1.771E-05 & 4.920E-01 \\
  75 & 3E+02 & 1.241E-07 & 3.479E-06 & 1.594E-05 & 2.280E-06 & 3.558E+00 & 4.968E-06 & 1.893E+00 & 9.115E-06 & 1.115E+00 & 1.470E-05 & 7.156E-01 & 2.141E-05 & 4.932E-01 \\
  75 & 1E+03 & 4.137E-07 & 1.158E-05 & 3.465E-05 & 3.057E-06 & 3.592E+00 & 6.606E-06 & 1.911E+00 & 1.203E-05 & 1.125E+00 & 1.928E-05 & 7.219E-01 & 2.794E-05 & 4.975E-01 \\
  75 & 3E+03 & 1.241E-06 & 3.474E-05 & 8.716E-05 & 4.248E-06 & 3.679E+00 & 9.071E-06 & 1.954E+00 & 1.636E-05 & 1.150E+00 & 2.599E-05 & 7.377E-01 & 3.\text{eff}E-05 & 5.082E-01 \\
  75 & 1E+04 & 4.135E-06 & 1.156E-04 & 2.591E-04 & 6.364E-06 & 3.854E+00 & 1.335E-05 & 2.043E+00 & 2.372E-05 & 1.201E+00 & 3.717E-05 & 7.695E-01 & 5.278E-05 & 5.299E-01 \\
  75 & 3E+04 & 1.240E-05 & 3.471E-04 & 7.321E-04 & 9.064E-06 & 3.997E+00 & 1.870E-05 & 2.115E+00 & 3.270E-05 & 1.242E+00 & 5.049E-05 & 7.953E-01 & 7.073E-05 & 5.473E-01 \\
  75 & 1E+05 & 4.136E-05 & 1.155E-03 & 2.379E-03 & 1.258E-05 & 3.910E+00 & 2.557E-05 & 2.066E+00 & 4.398E-05 & 1.213E+00 & 6.677E-05 & 7.762E-01 & 9.214E-05 & 5.341E-01 \\
  75 & 3E+05 & 1.241E-04 & 3.463E-03 & 7.089E-03 & 1.639E-05 & 3.873E+00 & 3.293E-05 & 2.046E+00 & 5.576E-05 & 1.201E+00 & 8.330E-05 & 7.685E-01 & 1.132E-04 & 5.288E-01 \\
  75 & 1E+06 & 4.136E-04 & 1.153E-02 & 2.361E-02 & 2.278E-05 & 3.889E+00 & 4.494E-05 & 2.055E+00 & 7.421E-05 & 1.206E+00 & 1.081E-04 & 7.720E-01 & 1.439E-04 & 5.312E-01 \\
  75 & 3E+06 & 1.240E-03 & 3.468E-02 & 7.111E-02 & 4.999E-05 & 3.928E+00 & 9.063E-05 & 2.075E+00 & 1.361E-04 & 1.218E+00 & 1.830E-04 & 7.799E-01 & 2.288E-04 & 5.367E-01 \\
  75 & 1E+07 & 4.137E-03 & 1.151E-01 & 2.355E-01 & 2.541E-04 & 3.910E+00 & 2.887E-04 & 2.068E+00 & 3.128E-04 & 1.214E+00 & 3.444E-04 & 7.777E-01 & 3.808E-04 & 5.353E-01 \\
100 & 1E+00 & 3.121E-10 & 2.816E-08 & 2.803E+00 & 1.861E-06 & 3.889E+00 & 4.125E-06 & 2.063E+00 & 7.649E-06 & 1.214E+00 & 1.242E-05 & 7.784E-01 & 1.820E-05 & 5.363E-01
\enddata
\tablecomments{Table 7 is published in its entirety in the electronic edition of the \textit{Astrophysical Journal Supplement}. A portion is shown here for guidance regarding its form and content.  $\kappa_{\text{R}}$ is the Rosseland Mean opacity and $\kappa_{\text{P}}$ is the Planck Mean opacity. Columns showing $T_{\rm eff}$ values of 3000-7000 K use this temperature, instead of the local temperature, in the weighting function.  The online version includes an additional significant figure for the density and opacities.}
\end{deluxetable}

\begin{deluxetable}{ccccccccccccccc}
\tabletypesize{\scriptsize}
\setlength{\tabcolsep}{0.01in} 
\tablecolumns{15}
\tablewidth{0pc}
\tablecaption{Mean Opacities for $\text{[M/H]}  = 1.7$}
\tablehead{
\colhead{} & \colhead{} & \colhead{}   &  \multicolumn{2}{c}{Local $T$} & \multicolumn{2}{c}{$T_{{\text{eff}}}=3000$ K} &  \multicolumn{2}{c}{$T_{\text{eff}}=4000$ K} &  \multicolumn{2}{c}{$T_{\text{eff}}=5000$ K} &  \multicolumn{2}{c}{$T_{\text{eff}}=6000$ K} &  \multicolumn{2}{c}{$T_{\text{eff}}=7000$ K} \\
%\cline{4-5} \cline{6-7} \cline{8-9} \cline{10-11} \cline{12-13} \cline{14-15} \\
\colhead{$T$} & \colhead{$P$}   & \colhead{$\rho$}  & \colhead{$\kappa_{\text{R}}$} & \colhead{$\kappa_{\text{P}}$} & \colhead{$\kappa_{\text{R}}$} & \colhead{$\kappa_{\text{P}}$} & \colhead{$\kappa_{\text{R}}$} & \colhead{$\kappa_{\text{P}}$} &  \colhead{$\kappa_{\text{R}}$} & \colhead{$\kappa_{\text{P}}$}  & \colhead{$\kappa_{\text{R}}$} & \colhead{$\kappa_{\text{P}}$} & \colhead{$\kappa_{\text{R}}$} & \colhead{$\kappa_{\text{P}}$} \\
\colhead{K} & \colhead{dyne cm$^{-2}$}   & \colhead{g cm$^{-3}$}  & \colhead{cm$^{2}$ g$^{-1}$} & \colhead{cm$^{2}$ g$^{-1}$} & \colhead{cm$^{2}$ g$^{-1}$} & \colhead{cm$^{2}$ g$^{-1}$} &  \colhead{cm$^{2}$ g$^{-1}$} & \colhead{cm$^{2}$ g$^{-1}$} & \colhead{cm$^{2}$ g$^{-1}$} & \colhead{cm$^{2}$ g$^{-1}$} & \colhead{cm$^{2}$ g$^{-1}$} & \colhead{cm$^{2}$ g$^{-1}$} & \colhead{cm$^{2}$ g$^{-1}$} & \colhead{cm$^{2}$ g$^{-1}$}} 
\startdata
  75 & 1E+00 & 4.383E-10 & 1.256E-08 & 3.413E-05 & 8.826E-07 & 5.454E+00 & 1.973E-06 & 2.903E+00 & 3.688E-06 & 1.710E+00 & 6.037E-06 & 1.097E+00 & 8.904E-06 & 7.561E-01 \\
  75 & 3E+00 & 1.314E-09 & 3.743E-08 & 1.868E-05 & 1.104E-06 & 5.454E+00 & 2.448E-06 & 2.903E+00 & 4.552E-06 & 1.710E+00 & 7.424E-06 & 1.097E+00 & 1.092E-05 & 7.561E-01 \\
  75 & 1E+01 & 4.383E-09 & 1.242E-07 & 1.351E-05 & 1.316E-06 & 5.455E+00 & 2.902E-06 & 2.903E+00 & 5.376E-06 & 1.710E+00 & 8.746E-06 & 1.097E+00 & 1.283E-05 & 7.562E-01 \\
  75 & 3E+01 & 1.314E-08 & 3.719E-07 & 1.269E-05 & 1.493E-06 & 5.457E+00 & 3.284E-06 & 2.904E+00 & 6.074E-06 & 1.711E+00 & 9.865E-06 & 1.097E+00 & 1.445E-05 & 7.564E-01 \\
  75 & 1E+02 & 4.383E-08 & 1.237E-06 & 1.463E-05 & 1.755E-06 & 5.462E+00 & 3.851E-06 & 2.907E+00 & 7.108E-06 & 1.712E+00 & 1.152E-05 & 1.098E+00 & 1.685E-05 & 7.571E-01 \\
  75 & 3E+02 & 1.314E-07 & 3.710E-06 & 2.103E-05 & 2.151E-06 & 5.478E+00 & 4.703E-06 & 2.915E+00 & 8.651E-06 & 1.717E+00 & 1.397E-05 & 1.101E+00 & 2.038E-05 & 7.590E-01 \\
  75 & 1E+03 & 4.383E-07 & 1.235E-05 & 4.208E-05 & 2.887E-06 & 5.530E+00 & 6.267E-06 & 2.941E+00 & 1.145E-05 & 1.732E+00 & 1.840E-05 & 1.111E+00 & 2.670E-05 & 7.656E-01 \\
  75 & 3E+03 & 1.314E-06 & 3.705E-05 & 1.005E-04 & 4.024E-06 & 5.663E+00 & 8.640E-06 & 3.009E+00 & 1.565E-05 & 1.771E+00 & 2.493E-05 & 1.135E+00 & 3.590E-05 & 7.821E-01 \\
  75 & 1E+04 & 4.383E-06 & 1.234E-04 & 2.871E-04 & 6.052E-06 & 5.956E+00 & 1.278E-05 & 3.157E+00 & 2.283E-05 & 1.856E+00 & 3.590E-05 & 1.189E+00 & 5.108E-05 & 8.186E-01 \\
  75 & 3E+04 & 1.314E-05 & 3.700E-04 & 7.894E-04 & 8.629E-06 & 6.150E+00 & 1.794E-05 & 3.254E+00 & 3.157E-05 & 1.911E+00 & 4.893E-05 & 1.223E+00 & 6.871E-05 & 8.420E-01 \\
  75 & 1E+05 & 4.383E-05 & 1.232E-03 & 2.542E-03 & 1.197E-05 & 6.023E+00 & 2.456E-05 & 3.182E+00 & 4.254E-05 & 1.868E+00 & 6.488E-05 & 1.195E+00 & 8.977E-05 & 8.224E-01 \\
  75 & 3E+05 & 1.314E-04 & 3.693E-03 & 7.551E-03 & 1.557E-05 & 5.968E+00 & 3.161E-05 & 3.153E+00 & 5.394E-05 & 1.850E+00 & 8.097E-05 & 1.184E+00 & 1.104E-04 & 8.146E-01 \\
  75 & 1E+06 & 4.383E-04 & 1.228E-02 & 2.508E-02 & 2.163E-05 & 5.970E+00 & 4.315E-05 & 3.154E+00 & 7.189E-05 & 1.851E+00 & 1.053E-04 & 1.184E+00 & 1.406E-04 & 8.150E-01 \\
  75 & 3E+06 & 1.314E-03 & 3.681E-02 & 7.517E-02 & 4.848E-05 & 5.976E+00 & 8.926E-05 & 3.158E+00 & 1.353E-04 & 1.853E+00 & 1.823E-04 & 1.186E+00 & 2.279E-04 & 8.161E-01 \\
  75 & 1E+07 & 4.383E-03 & 1.226E-01 & 2.505E-01 & 2.706E-04 & 6.000E+00 & 3.073E-04 & 3.171E+00 & 3.285E-04 & 1.862E+00 & 3.568E-04 & 1.192E+00 & 3.903E-04 & 8.202E-01 \\
100 & 1E+00 & 3.305E-10 & 2.884E-08 & 2.655E+00 & 1.800E-06 & 5.900E+00 & 3.995E-06 & 3.129E+00 & 7.417E-06 & 1.841E+00 & 1.206E-05 & 1.180E+00 & 1.767E-05 & 8.127E-01
\enddata
\tablecomments{Table 8 is published in its entirety in the electronic edition of the \textit{Astrophysical Journal Supplement}. A portion is shown here for guidance regarding its form and content.  $\kappa_{\text{R}}$ is the Rosseland Mean opacity and $\kappa_{\text{P}}$ is the Planck Mean opacity. Columns showing $T_{\rm eff}$ values of 3000-7000 K use this temperature, instead of the local temperature, in the weighting function.  The online version includes an additional significant figure for the density and opacities.}
\end{deluxetable}

\end{turnpage}

%\bibliographystyle{apj}
%\bibliography{references}

\end{document}